\documentclass{article}
\usepackage{amssymb}
\usepackage{amsmath}

\begin{document}
\centerline{\Large \bf The Chronological Operator Algebra}
\medskip
\centerline{\Large \bf and}
\medskip
\centerline{\Large \bf Formal Solutions of Differential Equations}

\vskip 2cm
\centerline{\sc Yu. N. Kosovtsov}
\medskip
\centerline{Lviv Radio Engineering Research Institute, Ukraine}
\centerline{email: {\tt kosovtsov@escort.lviv.net}} \vskip 1cm
\begin{abstract}
The aim of this paper is twofold. First, we obtain the explicit
exact formal solutions of differential equations of different types
in the form with Dyson chronological operator exponents. This allows
us to deal directly with the solutions to the equations rather than
the equations themselves. Second, we consider in detail the algebraic
properties of chronological operators, yielding an extensive family
of operator identities. The main advantage of the approach is to handle the
formal solutions at least as well as ordinary functions. We examine from a general standpoint
linear and non-linear ODEs of any order, systems of ODEs, linear
operator ODEs, linear PDEs and systems of linear PDEs for one
unknown function. The methods and techniques involved are demonstrated
on examples from important differential equations of mathematical
physics.
\end{abstract}

\section{Introduction}

There are several universal approaches to solving DE. The most
promising of them are Lie symmetry methods \cite {Lie}, \cite
{Olver} and advanced methods for the linear equations which are
based on differential Galois theory \cite {Singer}.

The aim of this contribution is to show examples of the
possibilities for solving differential equations with the method
based on Dyson's operator solution representation in the form of
time-ordered exponentials \cite {Dyson} (see also \cite {Bogolubov},
\cite {Kirznits}). It is well-known that Dyson's use of
time-ordering is the fundamental conceptual tool in quantum field
theory. This tool has now become a natural part of many branches of
physics and is even used in parts of engineering.

The method allows us to find the explicit exact operator solutions
of many problems if we succeed in reformulating a given problem into
a first-order linear one. The next step of the approach is directed
to transforming the operator solution into a more practical,
calculable or useful expression. The use of operator methods is
efficient in manipulating time-ordered expressions with help of
exponential identities like the Baker-Campbell-Hausdorff formulae
\cite {Magnus}, \cite {Wilcox}. Our second aim in this paper is to
give general methods for obtaining such identities.

We have in mind that at every stage the chronological exponential can
be expanded in formal operator power series neglecting all
the convergence problems that can appear with, for example, analytical functions.

As a result, this method allows us to deal directly with the solutions
to the equations rather than the equations themselves.

The main advantages of the method are in its compactness, clarity
and simplicity. It is also essential that we can handle the operator
expressions in manner similar to ordinary functions.

It is clear that there are some correlations between the operator method
and other common methods, especially the Lie approach.

We choose example equations for the operator
method from important differential equations from many branches
of science. The main aspects of the method are quite easily generalized
for more complicated equations.

As the paper is addressed to practical problems in many branches of
physics and engineering for which solutions are not yet at hand and to
avoid complications, we choose a simple (and non-rigorous in
places) manner of exposition in that we can always verify the
solutions by substitution into the initial equations. Analytical
restrictions are obvious from the context as a rule.

\section{Definitions and notations}

An \emph{operator} ${\bf A}$ is defined as a mapping of a function $f$ (from a ring $R$) into a function ${\bf A}f$ (from the same ring $R$). An operator ${\bf A}$ is \emph{linear} if it maps any two functions $\phi$ and $\psi$ in such a way that
\begin{equation}
{\bf A}(\phi + \psi ) = {\bf A} \phi + {\bf A} \psi.
\notag
\end{equation}
Operators will be written in the bold font seen above to avoid any ambiguities with the exception being the usual differential operator $\frac{\partial}{\partial x}$ notation. In every case throughout this paper, we shall be dealing exclusively with operators that are linear.

The operator ${\bf \Delta}$ is \emph {derivative} if
\begin{equation}
{\bf \Delta}({\bf A + B })f = {\bf \Delta A} f + {\bf \Delta B} f,
\notag
\end{equation}
\begin{equation}
{\bf \Delta}{\bf A}f = ({\bf \Delta A}) f + {\bf A}{\bf \Delta} f
\notag
\end{equation}
for any $f$ (from a ring $R$).
It follows from this definition that the \emph{inner derivative} of an operator ${\bf A}$ is
\begin{equation}
({\bf \Delta A}) = {\bf \Delta A}  - {\bf A\Delta}= [{\bf
\Delta},{\bf A}], \notag
\end{equation}
where $[{\bf \Delta},{\bf A}]$ is the conventional notation of the two
operator commutator. In this paper we will consider only \emph
{differentiable} operators ${\bf A}$ in sense that the commutator
$[{\bf \Delta},{\bf A}]$ exists.

If ${\bf \Delta}= \frac{\partial}{\partial x}$ then such a definition
of the inner derivative of the differentiable operator ${\bf A}(t)$ is in
accord with the classical analytic definition
\begin{equation}
\frac{\partial {\bf A}(x)}{\partial x} = \lim_{\epsilon \to 0} \frac{{\bf A}(x+\epsilon )-{\bf A}(x)}{\epsilon }.
\notag
\end{equation}

We will find need for an \emph{operator-valued function} $F({\bf A})$ which is an operator itself. Only functions which can be described by power series are considered in this work.

Important examples for this paper are the following types of linear operators:
\begin{equation}
{\bf L_1} = f_0 + f_1  \frac{\partial}{\partial x_1}+ ... + f_n \frac{\partial}{\partial x_n},
\label{L1}
\end{equation}
where $f_i$ are arbitrary functions of $t,x_1,...,x_n$, ${\bf L_1}$ is
obviously a derivative and
\begin{equation}
{\bf L_2} = F (t,x_1,...,x_n,\frac{\partial}{\partial x_1}, ... , \frac{\partial}{\partial x_n}).
\notag
\end{equation}
Though ${\bf L_2}$ is a generalization of ${\bf L_1}$, nevertheless
${\bf L_2}$ \emph{is not} a derivative in general, as e.g.,
$(\frac{\partial}{\partial x})^2=\frac{\partial^2}{\partial x^2}$
does not satisfy the above definition of derivative.

\section{The linear first-order differential equations and basic identities for chronological operator exponents}

\subsection{Dyson's form of the solution of first-order linear differential equation}

Let us start with the \emph{first-order linear} differential
equation with respect to $t$ for $u(t,\vec{\rho})=u(t)$ :
\begin{equation}
\frac{\partial u(t)}{\partial t} ={\bf L}(t)u(t)
\label{base}
\end{equation}
with the initial condition
\begin{equation}
u(t)|_{t=a}=v,
\label{basein}
\end{equation}
where $\vec{\rho}$ is a set of parameters say $x_1,...,x_n$, ${\bf
L}(t) ={\bf L}(t,\vec{\rho})$  is a \emph{linear operator} which
does not depend on $\frac{\partial}{\partial t}$ explicitly, and $v$
is some function of $\vec{\rho}$.

This equation can be solved by the following iteration scheme
\begin{equation}
u_0(t) = v,
\notag
\end{equation}
\begin{equation}
\frac{\partial u_n(t)}{\partial t} = {\bf L}(t)u_{n-1}(t),
\notag
\end{equation}
which for $t>a$ leads to
\begin{equation}
u(t)={\bf E}(t,a;{\bf L}(t))v,
\label{Ev}
\end{equation}
where operator {\bf E} has the following series expansion (${\bf 1}$ is the identity operator)
\begin{align}
{\bf E}(t,a;&{\bf L}(t))= {\bf 1}+\int_a^t d\tau\,{\bf L}(\tau) +
\notag \\ & \dots +\int_a^t d\tau_1\int_a^{\tau_1} d\tau_2\dots
\int_a^{\tau_{n-1}} d\tau_n\,\ {\bf L}(\tau_1)\,\ {\bf
L}(\tau_2)\dots {\bf L}(\tau_n)+ \dots \label{Eseries}
\end{align}

The solution (\ref{Ev}) (taking into account (\ref{Eseries})) has the following properties:

(a). Operator ${\bf E}$ satisfies the \emph{operator} differential equation
\begin{equation}
\frac{\partial {\bf E}}{\partial t} ={\bf L}(t){\bf E}
\label{dE}
\end{equation}
with initial condition
\begin{equation}
{\bf E}|_{t=a} = {\bf 1},
\label{dEin}
\end{equation}
or the equivalent integral equation
\begin{equation}
{\bf E}(t) ={\bf 1}+\int_a^t d\tau\,{\bf L}(\tau){\bf E}(\tau).
\notag
\end{equation}

(b). All operators {\bf L} in the series expansion of {\bf E} are
\emph {ordered} in sense that any operator term ${\bf L}(\tau_1)\,
{\bf L}(\tau_2)\dots {\bf L}(\tau_n)$ corresponds to the requirement
that $\tau_1\geq\tau_2\geq\dots\geq\tau_n$. The ordering here is
very essential as operators ${\bf L}(\tau_i)$ and ${\bf L}(\tau_j)$
do not commute in general.

The property (a) suggests that the operator {\bf E} is a kind of
$\exp \{\int_a^t d\tau\,{\bf L}(\tau)\}$ but expansion of the
exponent here does not lead to the required operator ordering (b),
as for example, the third term of the {\bf E} expansion
\begin{equation}
\int_a^t d\tau_1\int_a^{\tau_1} d\tau_2\,\, {\bf L}(\tau_1)\, {\bf L}(\tau_2)
\notag
\end{equation}
differs from the third term of the exponent expansion
\begin{equation}
\frac{1}{2}\int_a^t d\tau_1\int_a^t d\tau_2\, {\bf L}(\tau_1)\, {\bf L}(\tau_2).
\notag
\end{equation}

The next trick was proposed by Freeman J. Dyson \cite {Dyson} (see
also \cite {Bogolubov}, \cite {Kirznits}). Let us demonstrate it on
the third term of the {\bf E} expansion. By changing notation
of integration variables and integration order we obtain
\begin{equation}
\int_a^t d\tau_1\int_a^{\tau_1} d\tau_2\, {\bf L}(\tau_1)\,{\bf
L}(\tau_2)=\int_a^t d\tau_1\int_{\tau_1}^t d\tau_2\, {\bf
L}(\tau_2)\, {\bf L}(\tau_1), \notag
\end{equation}
so
\begin{equation}
\int_a^t d\tau_1\int_a^{\tau_1} d\tau_2\, {\bf L}(\tau_1)\, {\bf
L}(\tau_2)=\frac{1}{2}\int_a^t d\tau_1 \{ \int_a^{\tau_1} d\tau_2\,
{\bf L}(\tau_1)\, {\bf L}(\tau_2)+\int_{\tau_1}^t d\tau_2\, {\bf
L}(\tau_2)\, {\bf L}(\tau_1) \}. \notag
\end{equation}
If we now introduce a {\bf T}-operator defined as
\[ {\bf T} \{ {\bf
L}(\tau_1){\bf L}(\tau_2)) \}=\left \{
\begin{array}{rll}
{\bf L}(\tau_1){\bf L}(\tau_2))& \mbox{if} & \tau_1 \geq \tau_2;
\\
{\bf L}(\tau_2){\bf L}(\tau_1))& \mbox{if} & \tau_1 < \tau_2,
\end{array}\right.
\]
then $n$-th term of the right-hand side of (\ref{Eseries}) can be transformed to
\begin{align}
\int_a^t d\tau_1\int_a^{\tau_1} &d\tau_2\dots \int_a^{\tau_{n-1}}
d\tau_n\,\ {\bf L}(\tau_1)\,\ {\bf L}(\tau_2)\dots {\bf
L}(\tau_n)=\notag \\ &\frac{1}{n!}\int_a^t d\tau_1\int_a^t
d\tau_2\dots \int_a^t d\tau_n\,\ {\bf T} \{ {\bf L}(\tau_1)\,\ {\bf
L}(\tau_2)\dots {\bf L}(\tau_n) \}. \notag
\end{align}
Removing the {\bf T}-operator outside of the integral signs means we can
express
\begin{align}
{\bf E}(t,a;&{\bf L}(t))= {\bf T} \{ {\bf 1}+\int_a^t d\tau\,{\bf
L}(\tau) + \notag \\ & \dots +\frac{1}{n!}\int_a^t d\tau_1\int_a^t
d\tau_2\dots \int_a^t d\tau_n\,\ {\bf L}(\tau_1)\,\ {\bf
L}(\tau_2)\dots {\bf L}(\tau_n)+ \dots \}\notag
\end{align}
or in the final form with the \emph{chronological} operator exponential
\begin{equation}
{\bf E}(t,a;{\bf L}(t)) ={\bf T}\exp \{\int_a^t d\tau\,{\bf L}(\tau)\},
\label{ET}
\end{equation}
where we have in mind that (\ref{ET}) represents the series
expansion and that
\begin{align}
{\bf T}\, \{{\bf L}(\tau_1)\,\ {\bf L}(\tau_2)\dots {\bf L}(\tau_n) \}={\bf L}&(\tau_{\alpha_1})\,\ {\bf L}(\tau_{\alpha_2})\dots {\bf L}(\tau_{\alpha_n}) .\notag \\ & \tau_{\alpha_1}\geq \tau_{\alpha_2}\geq \dots \geq \tau_{\alpha_m}\notag
\end{align}

The expression (\ref{ET}) is the formal solution of the operator differential equation (\ref{dE}) with initial condition (\ref{dEin}).
The formal solution of the problem (\ref{base}), (\ref{basein}) is obviously
\begin{equation}
u(t) ={\bf T}\exp \{\int_a^t d\tau\,{\bf L}(\tau)\} \, v .
\label{uT}
\end{equation}
The main advantage of solutions in the chronological
exponential form is the explicit dependence of all parameters of the
problem and the relatively clear way to obtaining \emph {approximate}
non-formal solutions through series expansion
\begin{align}
&{\bf T}\exp \{\int_a^t d\tau\,{\bf L}(\tau)\}= {\bf 1}+\int_a^t
d\tau\,{\bf L}(\tau) +\notag \\ & \dots +\int_a^t
d\tau_1\int_a^{\tau_1} d\tau_2\dots \int_a^{\tau_{n-1}} d\tau_n\,\
{\bf L}(\tau_1)\,\ {\bf L}(\tau_2)\dots {\bf L}(\tau_n)+ \notag \\ &
\int_a^t d\tau_1\int_a^{\tau_1} d\tau_2\dots \int_a^{\tau_{n}}
d\tau_{n+1}\,\ {\bf L}(\tau_1)\,\ {\bf L}(\tau_2)\dots {\bf
L}(\tau_{n+1})\,{\bf T}\exp \{\int_a^{\tau_{n+1}} d\xi\,{\bf
L}(\xi)\}\,.\notag
\end{align}

Analogously, the solution of the problem (\ref{base}), (\ref{basein}) for $t<a$ is similar to (\ref{uT}), but instead of the operator ${\bf T}$ it is necessary to replace it with the operator ${\bf T}_0$, which represents a product of ${\bf L}$'s, in reverse order
\begin{align}
{\bf T}_0\, \{{\bf L}(\tau_1)\,\ {\bf L}(\tau_2)\dots {\bf L}(\tau_n) \}={\bf L}&(\tau_{\alpha_1})\,\ {\bf L}(\tau_{\alpha_2})\dots {\bf L}(\tau_{\alpha_n}) .\notag \\ & \tau_{\alpha_1}\leq \tau_{\alpha_2}\leq \dots \leq \tau_{\alpha_m}\notag
\end{align}

The problem (\ref{base}), (\ref{basein}) is deterministic in the sense
that if we start from an initial condition at $t=a$ we can find a unique
$u(t,\vec{\rho})$ for any $t$. If we now consider equation (\ref{base})
with a new initial condition $u(t,\vec{\rho})|_{t=b}=v_1(\vec{\rho})$
and recalculate $u(t,\vec{\rho})$ for $t\geq b$ we obtain (for
$a \leq b \leq t$)
\begin{equation}
u(t) ={\bf T}\exp \{\int_a^t d\tau\,{\bf L}(\tau)\} \, v ={\bf
T}\exp \{\int_b^t d\tau\,{\bf L}(\tau)\} \, {\bf T}\exp \{\int_a^b
d\tau\,{\bf L}(\tau)\} \, v \notag
\end{equation}
for any $v$. The operators ${\bf T}$ here and everywhere in this
paper act on operators of their exponentials only. That is, for $a
\leq b \leq t$ we have the following operator identity
\begin{equation}
{\bf T}\exp \{\int_a^t d\tau\,{\bf L}(\tau)\} ={\bf T}\exp
\{\int_b^t d\tau\,{\bf L}(\tau)\} \, {\bf T}\exp \{\int_a^b
d\tau\,{\bf L}(\tau)\}. \label{TT}
\end{equation}

Analogously, if on the second stage we recalculate $u(t,\vec{\rho})$
from $t$ to $a$ backward, we obtain that
\begin{equation}
v={\bf T}_0\exp \{\int_t^a d\tau\,{\bf L}(\tau)\}u(t) ={\bf T}_0\exp \{-\int_a^t d\tau\,{\bf L}(\tau)\} \, {\bf T}\exp \{\int_a^t d\tau\,{\bf L}(\tau)\}v
\notag
\end{equation}
for any $v$. Therefore
\begin{equation}
{\bf T}_0\exp \{-\int_a^t d\tau\,{\bf L}(\tau)\} \, {\bf T}\exp \{\int_a^t d\tau\,{\bf L}(\tau)\}={\bf 1}.
\label{TT0}
\end{equation}
So the operator
\begin{equation}
{\bf T}_0\exp \{-\int_a^t d\tau\,{\bf L}(\tau)\}
\notag
\end{equation}
is inverse to
\begin{equation}
{\bf T}\exp \{\int_a^t d\tau\,{\bf L}(\tau)\}
\notag
\end{equation}
and these operators commute.

It can be easily seen that the operator
\begin{equation}
{\bf E}^{-1}(t)={\bf T}_0\exp \{\int_a^t d\tau\,{\bf L}(\tau)\}
\label{T00}
\end{equation}
is the solution of the following operator differential equation
\begin{equation}
\frac{\partial {\bf E}^{-1}(t)}{\partial t} ={\bf E}^{-1}(t){\bf L}(t)
\label{dE1}
\end{equation}
with an initial condition
\begin{equation}
{\bf E}(t)|_{t=a} = {\bf 1} \label{dEin1}
\end{equation}
and that
\begin{equation}
{\bf E}^{-1}(t) ={\bf 1}+\int_a^t d\tau\,{\bf E}^{-1}(\tau){\bf L}(\tau)
\label{dE1ser}
\end{equation}
which leads to the following series expansion
\begin{align}
&{\bf E}^{-1}(t)={\bf T}_0\exp \{\int_a^t d\tau\,{\bf L}(\tau)\}=
{\bf 1}+\int_a^t d\tau\,{\bf L}(\tau) + \notag \\ & \dots +\int_a^t
d\tau_1\int_a^{\tau_1} d\tau_2\dots \int_a^{\tau_{n- 1}} d\tau_n\,\
{\bf L}(\tau_n)\,\ {\bf L}(\tau_{n-1})\dots {\bf L}(\tau_1)+\notag
\\ & \int_a^t d\tau_1\int_a^{\tau_1} d\tau_2\dots
\int_a^{\tau_{n}} d\tau_{n+1}{\bf T}_0\exp \{\int_a^{\tau_{n+1}}
d\xi\,{\bf L}(\xi)\}\,{\bf L}(\tau_{n+1}) {\bf L}(\tau_n)\dots {\bf
L}(\tau_1)\,. \label{E1series}
\end{align}

It follows from properties (\ref{dE1})-(\ref{E1series}) (or from
(\ref{TT})) that for $a \leq b \leq t$
\begin{equation}
{\bf T}_0\exp \{\int_a^t d\tau\,{\bf L}(\tau)\} ={\bf T}_0\exp
\{\int_a^b d\tau\,{\bf L}(\tau)\} \, {\bf T}_0\exp \{\int_b^t
d\tau\,{\bf L}(\tau)\}. \label{T0T0}
\end{equation}

With property (\ref{dE}) at hand one can easily verify that the
solution of the \emph{inhomogeneous} linear differential equation
\begin{equation}
\frac{\partial u(t)}{\partial t} ={\bf L}(t)u(t)+ \phi(t) \qquad
(u(t)|_{t=a}=v) \notag
\end{equation}
has the following operator form ($t>a$)
\begin{equation}
u(t) ={\bf T}\exp \{ \int_a^t d\tau\,{\bf L}(\tau)\} \,v+\int_a^t d\tau\,{\bf T}\exp \{\int_\tau^t d\xi\,{\bf L}(\xi)\} \phi(\tau).
\label{baseinsol}
\end{equation}

The operator technique in the form given above is immediately
suitable only for solving the linear problems. As it will be
demonstrated later, an analogous approach is fruitful for more
complicated problems. However, before touching such problems we need
to consider the algebra of chronological operator exponentials in more
details.

\subsection{Basic transformation identities for chronological operator exponents}

For any linear operator ${\bf L}(t)$ (not necessarily invertible) we can
form the chronological operator exponential (\emph{here and later on}
$t>a$)
\begin{equation}
{\bf T}\exp \{\int_a^t d\tau\,{\bf L}(\tau)\}, \notag
\end{equation}
which is always invertible and differentiable with respect to $t$.

Suppose we have an arbitrary linear invertible differentiable operator ${\bf A}(t)$. From the obvious identity
\begin{equation}
\frac{\partial {\bf A}(t)}{\partial t}=\{ \frac{\partial {\bf A}(t)}{\partial t} \, {\bf A}^{-1}(t) \} {\bf A}(t) \qquad ({\bf A}(t)|_{t=a}={\bf A}(a))
\notag
\end{equation}
we obtain
\begin{equation}
{\bf A}(t)={\bf T}\exp \{\int_a^t d\tau\,\frac{\partial {\bf A}(\tau)}{\partial \tau} \, {\bf A}^{-1}(\tau)\}{\bf A}(a).
\label{AE}
\end{equation}
It follows that the chronological operator exponential can represent the $t$-dependence of \emph {any} linear invertible differentiable operator, and each such operator corresponds to the following linear first-order differential equation
\begin{equation}
\frac{\partial u(t)}{\partial t}=\{ \frac{\partial {\bf A}(t)}{\partial t} \, {\bf A}^{-1}(t) \} u(t) \qquad (u(t)|_{t=a}=v)
\notag
\end{equation}
with the solution
\begin{equation}
u(t)={\bf A}(t){\bf A}^{-1}(a) \, v.
\notag
\end{equation}

As the product of any two of such operators is the linear
invertible differentiable operator (i.e., they form group) then the
set of all chronological operators forms group too.

If we invert (\ref{AE})
\begin{equation}
{\bf A}^{-1}(t)={\bf A}^{-1}(a){\bf T}_0\exp \{-\int_a^t d\tau\,\frac{\partial {\bf A}(\tau)}{\partial \tau} \, {\bf A}^{-1}(\tau)\}
\notag
\end{equation}
and replace ${\bf A}^{-1}(t)$ by ${\bf B}(t)$ we get
\begin{align}
{\bf B}(t)&={\bf B}(a){\bf T}_0\exp \{-\int_a^t
d\tau\,\frac{\partial {\bf B}^{-1}(\tau)}{\partial \tau} \, {\bf
B}(\tau)\}=\notag\\&{\bf B}(a){\bf T}_0\exp \{\int_a^t d\tau\,{\bf
B}^{-1}(\tau)\frac{\partial {\bf B}(\tau)}{\partial \tau} \} \notag
\end{align}
and conclude that any linear invertible differentiable operator can
be represented as an inverse ordered operator exponential too. Hence, the two
main forms of opposite ordered operator exponents can be expressed
through each other (see (\ref{T0T}), (\ref{TT1}) below).

Let us consider the following operator
\begin{equation}
{\bf K}(t)={\bf b}(t) \,{\bf T}\exp \{\int_a^t d\tau\,{\bf A}(\tau)\},
\notag
\end{equation}
where ${\bf b}(t)$ is a linear invertible differentiable operator and ${\bf A}(t)$ is a linear one.

Differentiating ${\bf K}$ with respect to $t$ we have
\begin{equation}
\frac{\partial {\bf K}(t)}{\partial t} =\{ \frac{\partial {\bf b}(t)}{\partial t} {\bf b}^{-1}(t)+{\bf b}(t){\bf A}(t){\bf b}^{-1}(t) \} \, {\bf K}(t)
\label{bA}
\end{equation}
and
\begin{equation}
{\bf K}(t)|_{t=a}={\bf b}(a).
\label{bAin}
\end{equation}
As it was shown in the previous section, the operator differential equation (\ref{bA}) with an initial condition (\ref{bAin}) has a solution in the following form
\begin{equation}
{\bf K}(t)={\bf T}\exp \{\int_a^t d\tau\,[\frac{\partial {\bf
b}(\tau)}{\partial \tau} {\bf b}^{-1}(\tau)+{\bf b}(\tau){\bf
A}(\tau){\bf b}^{-1}(\tau)]\}\, {\bf b}(a),\notag
\end{equation}
so we obtain the important identity
\begin{equation}
{\bf b}(t) \,{\bf T}\exp \{\int_a^t d\tau\,{\bf A}(\tau)\}={\bf T}\exp \{\int_a^t d\tau\,[\frac{\partial {\bf b}(\tau)}{\partial \tau} {\bf b}^{-1}(\tau)+{\bf b}(\tau){\bf A}(\tau){\bf b}^{-1}(\tau)]\} \, {\bf b}(a).
\label{bprA}
\end{equation}

Now if
\begin{equation}
{\bf b}(t)={\bf T}\exp \{\int_a^t d\tau\,{\bf B}(\tau)\},
\notag
\end{equation}
then from (\ref{bprA}) we obtain one of the basic chronological
operator identities
\begin{align}
&{\bf T}\exp \{\int_a^t d\tau\,{\bf B}(\tau)\} \,{\bf T}\exp
\{\int_a^t d\tau\,{\bf A}(\tau)\}=\notag \\ &{\bf T}\exp \{\int_a^t
d\tau\,[{\bf B}(\tau)+{\bf T}\exp \{\int_a^\tau d\xi\,{\bf
B}(\xi)\}\,{\bf A}\,(\tau){\bf T}_0\exp \{-\int_a^\tau d\xi\,{\bf
B}(\xi)\}] \}. \label{BprA}
\end{align}

If we denote
\begin{equation}
{\bf C}(\tau)={\bf T}\exp \{\int_a^\tau d\xi\,{\bf B}(\xi)\}\,{\bf A}(\tau)\,{\bf T}_0\exp \{-\int_a^\tau d\xi\,{\bf B}(\xi)\}
\notag
\end{equation}
and solve it with respect to ${\bf A}$
\begin{equation}
{\bf A}(\tau)={\bf T}_0\exp \{-\int_a^\tau d\xi\,{\bf B}(\xi)\}\,{\bf C}(\tau)\,{\bf T}\exp \{\int_a^\tau d\xi\,{\bf B}(\xi)\},
\notag
\end{equation}
then it follows immediately from (\ref{BprA}) that
\begin{align}
{\bf T}\exp & \{\int_a^t d\tau\,[{\bf B}(\tau)+{\bf C}(\tau)]\}
={\bf T}\exp \{\int_a^t d\tau\,{\bf B}(\tau)\} \times \notag \\ &
{\bf T}\exp \{\int_a^t d\tau\,{\bf T}_0\exp \{-\int_a^\tau
d\xi\,{\bf B}(\xi)\}\,{\bf C}(\tau)\,{\bf T}\exp \{\int_a^\tau
d\xi\,{\bf B}(\xi)\} \}. \label{BsumA}
\end{align}

Identities (\ref{BprA}) and (\ref{BsumA}) are generalizations of the
well-known Baker-Campbell-Hausdorff (BCH) and Zassenhaus formulae
for \emph{``$t$-dependent"} operators in the sense that the classical BCH
formula merges two exponential operators into a single one and
Zassenhaus formula splits an exponential operator into a product of
exponential operators.

At first sight the identities (\ref{BprA}) and (\ref{BsumA})
represent a circle of some kind. To demonstrate the value of
these identities we first consider their classical expansions and
later (in Subsection 5.2 below) we show that in many cases of practical
important that involve (\ref{BprA}) and (\ref{BsumA})
operator expressions can be obtained exactly by a more direct way
without using the expansions like (\ref{BCHii}), (\ref{BCHii2}).

\subsection{Baker-Campbell-Hausdorff and Zassenhaus formulae for chronological operator exponents}

Let us consider the following construction
\begin{equation}
{\bf K}(a)={\bf T}\exp \{\int_a^t d\tau\,{\bf B}(\tau)\} \,{\bf
A}(t) \,{\bf T}_0\exp \{-\int_a^t d\tau\,{\bf B}(\tau)\}. \label{Ka}
\end{equation}
If we differentiate ${\bf K}(a)$ with respect to $a$ we obviously obtain the following expression
\begin{equation}
\frac{\partial {\bf K}(a)}{\partial a}={\bf T}\exp \{\int_a^t d\tau\,{\bf B}(\tau)\} \,[-{\bf B}(a){\bf A}(t)+{\bf A}(t){\bf B}(a)] \,{\bf T}_0\exp \{-\int_a^t d\tau\,{\bf B}(\tau)\},
\notag
\end{equation}
where
\begin{equation}
{\bf K}(a)|_{a=t}={\bf A}(t).
\notag
\end{equation}
If we use the commutator notation ${\bf A}(t){\bf B}(a)-{\bf B}(a){\bf A}(t)=[{\bf A}(t),{\bf B}(a)]$ we get after integration that
\begin{equation}
{\bf K}(a)={\bf A}(t)+\int_t^ad\tau \,{\bf T}\exp \{\int_\tau^t d\xi\,{\bf B}(\xi)\} \,[{\bf A}(t),{\bf B}(\tau)] \,{\bf T}_0\exp \{-\int_\tau^t d\xi\,{\bf B}(\xi)\}
\notag
\end{equation}
or
\begin{align}
&{\bf T}\exp \{\int_a^t d\tau\,{\bf B}(\tau)\} \,{\bf A}(t) \,{\bf
T}_0\exp \{-\int_a^t d\tau\,{\bf B}(\tau)\}=\notag \\ & {\bf
A}(t)-\int_a^t d\tau\,{\bf T}\exp \{\int_\tau^t d \xi \,{\bf
B}(\xi)\} \,[{\bf A}(t),{\bf B}(\tau)] \,{\bf T}_0\exp
\{-\int_\tau^t d\xi\,{\bf B}(\xi)\}. \label{BCHi}
\end{align}
This is an analog of so-called integral BCH formula. Iterations of
(\ref{BCHi}) leads to the following expansion
\begin{align}
&{\bf T}\exp \{\int_a^t d\tau\,{\bf B}(\tau)\} \,{\bf A}(t) \,{\bf
T}_0\exp \{-\int_a^t d\tau\,{\bf B}(\tau)\}=\notag \\ & {\bf
A}(t)-\int_a^t d\tau\,[{\bf A}(t),{\bf B}(\tau)] +\int_a^t
d\tau_1\int_{\tau_1}^t d\tau_2\,[[{\bf A}(t),{\bf B}(\tau_1)],{\bf
B}(\tau_2)] -\dots\notag \\ &+(-1)^n \int_a^t d\tau_1
\int_{\tau_1}^t d\tau_2 \dots  \int_{\tau_{n-1}}^t d\tau_n  \,{\bf
T}\exp \{\int_{\tau_n}^t d \xi \,{\bf B}(\xi)\}\times \notag \\
&[[...[{\bf A}(t),{\bf B}(\tau_1)],{\bf B}(\tau_2)]...],{\bf
B}(\tau_n)] \,{\bf T}_0\exp \{-\int_{\tau_n}^t d\xi\,{\bf B}(\xi)\}
\label{BCHii}.
\end{align}
If we now substitute (\ref{BCHii}) into (\ref{BprA}) we obtain the more conventional form of BCH formula
\begin{align}
&{\bf T}\exp \{\int_a^t d\tau\,{\bf B}(\tau)\} \,{\bf T}\exp
\{\int_a^t d\tau\,{\bf A}(\tau)\}=\notag \\ &{\bf T}\exp \{\int_a^t
d\tau\,\{ {\bf A}(\tau)+{\bf B}(\tau)-\int_a^\tau d\tau_1\,[{\bf
A}(\tau),{\bf B}(\tau_1)] +\notag \\ & \int_a^\tau
d\tau_1\int_{\tau_1}^\tau d\tau_2\,[[{\bf A}(\tau),{\bf
B}(\tau_1)],{\bf B}(\tau_2)] -\dots\notag \\ & +(-1)^n \int_a^\tau
d\tau_1 \int_{\tau_1}^\tau d\tau_2 \dots  \int_{\tau_{n-1}}^\tau
d\tau_n  \,{\bf T}\exp \{\int_{\tau_n}^\tau d \xi \,{\bf
B}(\xi)\}\times \notag \\ &[[...[{\bf A}(\tau),{\bf B}(\tau_1)],{\bf
B}(\tau_2)]...],{\bf B}(\tau_n)] \,{\bf T}_0\exp
\{-\int_{\tau_n}^\tau d\xi\,{\bf B}(\xi)\}\} \}. \label{BprABCH}
\end{align}
The importance of the last BCH formula (\ref{BprABCH}) lies not in the details of the formula, but in the fact that there is one, and the fact that it gives the product of two operators
\begin{equation}
{\bf T}\exp \{\int_a^t d\tau\,{\bf B}(\tau)\} \,{\bf T}\exp \{\int_a^t d\tau\,{\bf A}(\tau)\}
\notag
\end{equation}
as one operator in terms of ${\bf A}$ and ${\bf B}$, brackets of
${\bf A}$ and ${\bf B}$, brackets of brackets, etc.

Analogously, starting from
\begin{equation}
{\bf K}(\lambda)={\bf T}_0\exp \{-\int_a^\lambda d\tau\,{\bf
B}(\tau)\} \,{\bf A}(t) \,{\bf T}\exp \{\int_a^\lambda d\tau\,{\bf
B}(\tau)\} \label{BCH0}
\end{equation}
and differentiating ${\bf K}(\lambda)$ with respect to $\lambda$ we obtain the following expression
\begin{equation}
\frac{\partial {\bf K}(\lambda)}{\partial \lambda}={\bf T}_0\exp \{-\int_a^\lambda  d\tau\,{\bf B}(\tau)\} \,[-{\bf B}(\lambda){\bf A}(t)+{\bf A}(t){\bf B}(\lambda)] \,{\bf T}\exp \{\int_a^\lambda  d\tau\,{\bf B}(\tau)\},
\notag
\end{equation}
where
\begin{equation}
{\bf K}(\lambda)|_{\lambda=a}={\bf A}(t).
\notag
\end{equation}
After integration we get that
\begin{equation}
{\bf K}(\lambda)={\bf A}(t)+\int_a^\lambda d\tau \,{\bf T}_0\exp \{-\int_a^\tau d\xi\,{\bf B}(\xi)\} \,[{\bf A}(t),{\bf B}(\tau)] \,{\bf T}\exp \{\int_a^\tau d\xi\,{\bf B}(\xi)\}
\notag
\end{equation}
for any $\lambda$. By setting $\lambda=t$ we arrive to
\begin{align}
&{\bf T}_0\exp \{-\int_a^t d\tau\,{\bf B}(\tau)\} \,{\bf A}(t)
\,{\bf T}\exp \{\int_a^t d\tau\,{\bf B}(\tau)\}=\notag \\ & {\bf
A}(t)+\int_a^t d\tau\,{\bf T}_0\exp \{-\int_a^\tau d \xi \,{\bf
B}(\xi)\} \,[{\bf A}(t),{\bf B}(\tau)] \,{\bf T}\exp \{\int_a^\tau
d\xi\,{\bf B}(\xi)\}. \label{BCHi2}
\end{align}
This is a mirror-like twin brother of the integral BCH formula. We
note here that (\ref{BCHi2}) is the corollary of (\ref{BCHi}) if we
have in mind the properties of the chronological operators (\ref{TT}),
(\ref{TT0}) and (\ref{T0T0}). Iterations of (\ref{BCHi2}) lead to
the following expansion (compare with (\ref{BCHii}))
\begin{align}
&{\bf T}_0\exp \{-\int_a^t d\tau\,{\bf B}(\tau)\} \,{\bf A}(t) \,{\bf T}\exp \{\int_a^t d\tau\,{\bf B}(\tau)\}=\notag \\
& {\bf A}(t)+\int_a^t d\tau\,[{\bf A}(t),{\bf B}(\tau)]+\int_a^t d\tau_1\int_a^{\tau_1} d\tau_2\,[[{\bf A}(t),{\bf B}(\tau_1)],{\bf B}(\tau_2)] + \notag \\
&\hspace{7 mm} \int_a^t d\tau_1 \int_a^{\tau_1} d\tau_2 \dots  \int_a^{\tau_{n-1}} d\tau_n  \,{\bf T}_0\exp \{-\int_a^{\tau_n} d \xi \,{\bf B}(\xi)\} \times \notag \\
&\hspace{7 mm} [[...[{\bf A}(t),{\bf B}(\tau_1)],{\bf B}(\tau_2)]...],{\bf B}(\tau_n)] \,{\bf T}\exp \{\int_a^{\tau_n} d\xi\,{\bf B}(\xi)\}.
\label{BCHii2}
\end{align}
If we now substitute (\ref{BCHi2}) into (\ref{BsumA}) we obtain
\begin{align}
{\bf T}\exp & \{ \int_a^t d\tau \,[{\bf B}(\tau)+{\bf C}(\tau)]\}
={\bf T}\exp \{ \int_a^t d \tau\,{\bf B}(\tau)\} {\bf T}\exp \{
\int_a^t d\tau  \, [{\bf C}(\tau) + \notag \\ & \int_a^{\tau} d
\tau_1 {\bf T}_0 \exp \{-\int_a^{\tau_1} d\xi \,{\bf B}(\xi)\} \,
[{\bf C}(\tau),{\bf B}(\tau_1)] \,{\bf T}\exp \{\int_a^{\tau_1}
d\xi\,{\bf B}(\xi)\} ] \}
\end{align}
and with the help of the original identity (\ref{BsumA}) we further obtain
\begin{align}
{\bf T}\exp & \{ \int_a^t d\tau \,[{\bf B}(\tau)+{\bf C}(\tau)]\}
={\bf T}\exp \{ \int_a^t d \tau\,{\bf B}(\tau)\} \, \,  {\bf T}\exp
\{ \int_a^t d \tau\,{\bf C}(\tau)\}\times \notag \\ & {\bf T}\exp \{
\int_a^t d\tau  \, \int_a^{\tau} d \tau_1 \, {\bf T}_0 \exp
\{-\int_a^{\tau} d\xi \,{\bf C}(\xi)\} \, {\bf T}_0 \exp
\{-\int_a^{\tau_1} d\xi \,{\bf B}(\xi)\} \times \notag \\ & [{\bf
C}(\tau),{\bf B}(\tau_1)] \,  \,{\bf T}\exp \{\int_a^{\tau_1}
d\xi\,{\bf B}(\xi)\} \, {\bf T} \exp \{\int_a^{\tau} d\xi \,{\bf
C}(\xi)\} \, \}.\notag
\end{align}

Iterations of (\ref{BCHii2}) with the help of (\ref{BprA}) and (\ref{BsumA}) lead to more conventional form (but noticeably bulky than (\ref{BprABCH}), so we do not represent it here) of a generalized Zassenhaus formula, which nevertheless tells us that the operator
\begin{equation}
{\bf T}\exp \{ \int_a^t d\tau \,[{\bf B}(\tau)+{\bf C}(\tau)]\}
\notag
\end{equation}
can be expressed as product of operators in terms of ${\bf A}$ and
${\bf B}$, brackets of ${\bf A}$ and ${\bf B}$, brackets of
brackets, etc.

\subsection{Relationship between opposite ordered chronological exponents and generalized linear operator differential equations}
Let us consider now
\begin{align}
{\bf T}\exp & \{\int_a^t d\tau\,[{\bf A}(\tau)-{\bf A}(\tau)]\}
={\bf 1}={\bf T}\exp \{\int_a^t d\tau\,{\bf A}(\tau)\} \times \notag
\\ & {\bf T}\exp \{-\int_a^t d\tau\,{\bf T}_0\exp
\{-\int_a^\tau d\xi\,{\bf A}(\xi)\}{\bf A}(\tau){\bf T}\exp
\{\int_a^\tau d\xi\,{\bf A}(\xi)\} \}, \notag
\end{align}
so we deduce at once that
\begin{align}
{\bf T}&_0\exp \{-\int_a^t d\tau\,{\bf A}(\tau)\} = \notag \\ & {\bf
T}\exp \{-\int_a^t d\tau\,{\bf T}_0\exp \{-\int_a^\tau d\xi\,{\bf
A}(\xi)\}{\bf A}(\tau){\bf T}\exp \{\int_a^\tau d\xi\,{\bf A}(\xi)\}
\} \label{T0T}
\end{align}
and further
\begin{align}
{\bf T}&\exp \{\int_a^t d\tau\,{\bf A}(\tau)\} = \notag \\ & {\bf
T}_0\exp \{\int_a^t d\tau\,{\bf T}_0\exp \{-\int_a^\tau d\xi\,{\bf
A}(\xi)\}{\bf A}(\tau){\bf T}\exp \{\int_a^\tau d\xi\,{\bf A}(\xi)\}
\}. \label{TT1}
\end{align}

These identities allow us to solve the following equation
\begin{equation}
{\bf T}_0\exp\{\int_a^t d\xi\,{\bf A}(\xi)\}{\bf A}(t){\bf T}\exp\{-\int_a^t d\xi\,{\bf A}(\xi)\}={\bf B}(t)
\label{ab}
\end{equation}
with respect to ${\bf A}(t)$.
The solution of (\ref{ab}) is as follows
\begin{equation}
{\bf A}(t)={\bf T}_0\exp\{-\int_a^t d\xi\,{\bf B}(\xi)\}{\bf B}(t){\bf T}\exp\{\int_a^t d\xi\,{\bf B}(\xi)\}.
\label{ba}
\end{equation}
One can prove this by substitution of (\ref{ba}) into (\ref{ab}).
Really, in view of (\ref{T0T}) and (\ref{TT1}) we have the following
chain starting from (\ref{ab}) and then substituting (\ref{ba})
\begin{align}
&{\bf T}_0\exp\{\int_a^t d\xi\,{\bf A}(\xi)\}{\bf A}(t){\bf T}\exp\{-\int_a^t d\xi\,{\bf A}(\xi)\}=\notag \\
&{\bf T}_0\exp \{\int_a^t d\tau\,{\bf T}_0\exp \{-\int_a^\tau
d\xi\,{\bf B}(\xi)\}{\bf B}(\tau){\bf T}\exp \{\int_a^\tau
d\xi\,{\bf B}(\xi)\} \}\times \notag \\ & {\bf T}_0\exp\{-\int_a^t
d\xi\,{\bf B}(\xi)\}\,\,{\bf B}(t)\,\,{\bf T}\exp\{\int_a^t
d\xi\,{\bf B}(\xi)\}\times \notag \\ & {\bf T}\exp \{-\int_a^t
d\tau\,{\bf T}_0\exp \{-\int_a^\tau d\xi\,{\bf B}(\xi)\}{\bf
B}(\tau){\bf T}\exp \{\int_a^\tau d\xi\,{\bf B}(\xi)\} =\notag \\ &
{\bf T}\exp\{\int_a^t d\tau\,{\bf B}(\tau)\}{\bf T}_0\exp\{-\int_a^t
d\tau\,{\bf B}(\tau)\}\,\,{\bf B}(t)\,\,\times \notag \\ & {\bf
T}\exp\{\int_a^t d\tau\,{\bf B}(\tau)\}{\bf T}_0\exp\{-\int_a^t
d\tau\,{\bf B}(\tau)\}={\bf B}(t). \notag
\end{align}

Let us consider the following operator
\begin{equation}
{\bf K}(t)={\bf T}_0\exp \{\int_a^t d\tau\,{\bf A}(\tau)\} \,{\bf B}(t) \,{\bf T}\exp \{\int_a^t d\tau\,{\bf C}(\tau)\}.
\label{K0}
\end{equation}
Differentiating ${\bf K}(t)$ with respect to $t$ we have
\begin{align}
&\frac{\partial {\bf K}(t)}{\partial t}=\notag \\ &{\bf T}_0\exp \{\int_a^t d\tau\,{\bf A}(\tau)\} \,[{\bf A}(t){\bf B}(t)+\frac{\partial {\bf B}(t)}{\partial t}+{\bf B}(t){\bf C}(t)] \,{\bf T}\exp \{\int_a^t d\tau\,{\bf C}(\tau)\}
\notag
\end{align}
and obtain the following differential expression
\begin{equation}
\frac{\partial {\bf K}(t)}{\partial t}={\bf a}(t){\bf K}(t)+{\bf K}(t){\bf c}(t)+{\bf b}(t),
\label{KK}
\end{equation}
where
\begin{equation}
{\bf a}(t)={\bf T}_0\exp\{\int_a^t d\tau\,{\bf A}(\tau)\}\,\,{\bf A}(t)\,\,{\bf T}\exp\{-\int_a^t d\tau\,{\bf A}(\tau\},
\label{aA}
\end{equation}
\begin{equation}
{\bf c}(t)={\bf T}_0\exp\{-\int_a^t d\tau\,{\bf C}(\tau)\}\,\,{\bf C}(t)\,\,{\bf T}\exp\{\int_a^t d\tau\,{\bf C}(\tau)\},
\label{cC}
\end{equation}
\begin{equation}
{\bf b}(t)={\bf T}_0\exp\{\int_a^t d\tau\,{\bf A}(\tau)\}\,\,\frac{\partial {\bf B}(t)}{\partial t}\,\,{\bf T}\exp\{-\int_a^t d\tau\,{\bf C}(\tau\},
\label{bB}
\end{equation}
and moreover
\begin{equation}
{\bf K}(t)|_{t=a}={\bf B}(a).
\label{KBi}
\end{equation}

We can solve (\ref{aA}) and (\ref{cC}) with respect to ${\bf A}(t)$ and ${\bf C}(t)$ respectively
\begin{equation}
{\bf A}(t)={\bf T}_0\exp\{-\int_a^t d\tau\,{\bf a}(\tau)\}\,\,{\bf a}(t)\,\,{\bf T}\exp\{\int_a^t d\tau\,{\bf a}(\tau\},
\label{Aa}
\end{equation}
\begin{equation}
{\bf C}(t)=-{\bf T}_0\exp\{\int_a^t d\tau\,{\bf c}(\tau)\}\,\,{\bf c}(t)\,\,{\bf T}\exp\{-\int_a^t d\tau\,{\bf c}(\tau)\}.
\label{Cc}
\end{equation}
From (\ref{bB}) we have
\begin{equation}
\frac{\partial {\bf B}(t)}{\partial t}={\bf T}\exp \{-\int_a^t d\tau\,{\bf A}(\tau)\} \,\,{\bf b}(t) \,\,{\bf T}_0\exp \{-\int_a^t d\tau\,{\bf C}(\tau)\}
\notag
\end{equation}
and with (\ref{KBi}) it leads to
\begin{equation}
 {\bf B}(t)={\bf K}(a)+\int_a^t d\tau{\bf T}\exp \{-\int_a^\tau d\tau_1\,{\bf A}(\tau_1)\} \,\,{\bf b}(\tau) \,\,{\bf T}_0\exp \{-\int_a^\tau d\tau_1\,{\bf C}(\tau_1)\}
\notag
\end{equation}
and substituting (\ref{Aa}) and (\ref{Cc}) we have, taking into account identities (\ref{T0T}) and (\ref{TT1}), that
\begin{equation}
{\bf B}(t)={\bf K}(a)+\int_a^t d\tau \,{\bf T}_0\exp \{-\int_a^\tau
d\xi\,{\bf a}(\xi)\} \,\,{\bf b}(\tau)\, \,{\bf T}\exp
\{-\int_a^\tau d\xi\,{\bf c}(\xi)\} \notag
\end{equation}

If we now express ${\bf A}$, ${\bf B}$ and ${\bf C}$ in (\ref{K0}) via
${\bf a}$, ${\bf b}$ and ${\bf c}$ then it follows that the formal
solution of the operator differential equation (\ref{KK}) with
the initial condition ${\bf K}(t)|_{t=a}={\bf K}(a)$, where ${\bf a}$,
${\bf b}$ and ${\bf c}$ are any linear (not necessarily invertible)
operators, which do not depend on $\frac{\partial}{\partial t}$
explicitly, has the following form
\begin{align}
& {\bf K}(t)={\bf T}_0\exp\{\int_a^t d\tau\,{\bf
T}_0\exp\{-\int_a^\tau d\tau_1\,{\bf a}(\tau_1)\}\,\,{\bf
a}(\tau)\,\,{\bf T}\exp\{\int_a^\tau d\tau_1\,{\bf
a}(\tau_1\}\}\times\notag \\ &[{\bf K}(a)+\int_a^t d\tau \,{\bf
T}_0\exp \{-\int_a^\tau d\xi\,{\bf a}(\xi)\} \,\,{\bf b}(\tau)\,
\,{\bf T}\exp \{-\int_a^\tau d\xi\,{\bf c}(\xi)\}]\times\notag \\
&\notag {\bf T}\exp\{-\int_a^t d\tau\,{\bf T}_0\exp\{\int_a^\tau
d\tau_1\,{\bf c}(\tau_1)\}\,\,{\bf c}(\tau)\,\,{\bf
T}\exp\{-\int_a^\tau d\tau_1\,{\bf c}(\tau_1\}\}. \notag
\end{align}
By using identities (\ref{T0T}) and (\ref{TT1}) we arrive finally to
\begin{align}
&{\bf K}(t)={\bf T}\exp \{\int_a^t d\tau\,{\bf a}(\tau)\}
\,\times\notag \\ &[{\bf K}(a)+\int_a^t d\tau \,{\bf T}_0\exp
\{-\int_a^\tau d\xi\,{\bf a}(\xi)\} \,\,{\bf b}(\tau)\, \,{\bf
T}\exp \{-\int_a^\tau d\xi\,{\bf c}(\xi)\}] \,\times\notag \\
&{\bf T}_0\exp \{\int_a^t d\tau\,{\bf c}(\tau)\}, \label{opDEsol}
\end{align}
as the formal solution of (\ref{KK}), which can be verified directly
by substitution of (\ref{opDEsol}) into (\ref{KK}).

The equation (\ref{KK}) is a generalization of equations (\ref{dE}) and (\ref{dE1}). As we can easily see, their solutions (\ref{ET}) and (\ref{T00}) are particular cases of (\ref{opDEsol}).

\subsection{Differentiation of the chronological exponential with respect to a parameter}

In the previous sections we considered some analytical properties
of chronological operator exponentials mainly under differentiation on
$t$. It is very important from an analytical point of view to consider
differentiation of operator exponentials with respect to some parameter,
say $\alpha$.

Let us consider the following operator
\begin{equation}
\frac{\partial }{\partial \alpha}\,\,\,{\bf T}\exp \{\int_a^t d\tau\,{\bf A}(\tau,\alpha)\}
\notag
\end{equation}
and denote here
\begin{equation}
{\bf a}(t,\alpha)=\frac{\partial {\bf A}(t,\alpha)}{\partial
\alpha}=[\frac{\partial }{\partial \alpha},{\bf A}(t,\alpha)].
\notag
\end{equation}
From the identity (\ref{BCHi}) we have
\begin{align}
&{\bf T}\exp \{\int_a^t d\tau\,{\bf A}(\tau,\alpha)\} \,\,\,
\frac{\partial }{\partial \alpha}\,\,\,{\bf T}_0\exp \{-\int_a^t
d\tau\,{\bf A}(\tau,\alpha)\}=\notag \\ & \frac{\partial }{\partial
\alpha}-\int_a^t d\tau\,{\bf T}\exp \{\int_\tau^t d \xi \,{\bf
A}(\xi,\alpha)\} \,\,{\bf a}(\tau,\alpha) \,\,{\bf T}_0\exp
\{-\int_\tau^t d\xi\,{\bf A}(\xi,\alpha)\}, \notag
\end{align}
so
\begin{align}
&\frac{\partial }{\partial \alpha}\,\,\,{\bf T}_0\exp \{-\int_a^t
d\tau\,{\bf A}(\tau,\alpha)\}={\bf T}_0\exp \{-\int_a^t d\tau\,{\bf
A}(\tau,\alpha)\}\,[\frac{\partial }{\partial \alpha}\,-\notag \\ &
\int_a^t d\tau\,{\bf T}\exp \{\int_\tau^t d \xi \,{\bf
A}(\xi,\alpha)\} \,\,{\bf a}(\tau,\alpha) \,\,{\bf T}_0\exp
\{-\int_\tau^t d\xi\,{\bf A}(\xi,\alpha)\}]. \notag
\end{align}

Analogously from identity (\ref{BCHi2}) we have
\begin{align}
&{\bf T}_0\exp \{-\int_a^t d\tau\,{\bf A}(\tau,\alpha)\}
\,\,\,\frac{\partial }{\partial \alpha}\,\,\,{\bf T}\exp \{\int_a^t
d\tau\,{\bf A}(\tau,\alpha)\}=\notag \\ & \frac{\partial }{\partial
\alpha}+\int_a^t d\tau\,{\bf T}_0\exp \{-\int_a^\tau d \xi \,{\bf
A}(\xi,\alpha)\} \,\,{\bf a}(\tau,\alpha) \,\,{\bf T}\exp
\{\int_a^\tau d\xi\,{\bf A}(\xi,\alpha)\}, \notag
\end{align}
which leads to
\begin{align}
&\frac{\partial }{\partial \alpha}\,\,\,{\bf T}\exp \{\int_a^t
d\tau\,{\bf A}(\tau,\alpha)\}={\bf T}\exp \{\int_a^t d\tau\,{\bf
A}(\tau,\alpha)\}\,[\frac{\partial }{\partial \alpha}\,+\notag \\ &
\int_a^t d\tau\,{\bf T}_0\exp \{-\int_a^\tau d \xi \,{\bf
A}(\xi,\alpha)\} \,\,{\bf a}(\tau,\alpha) \,\,{\bf T}\exp
\{\int_a^\tau d\xi\,{\bf A}(\xi,\alpha)\}]. \label{DT}
\end{align}

\subsection{Shift operators}

To transform formal solutions into ordinary expressions we have to
express explicitly the action of the corresponding linear operators.
Every solved equation gives us an example of definite action of
a given linear operator. Thus, the list of ``good" operators is nonempty.
Our goal here is to study the properties of a certain class of
operators that we will refer to as shift operators.

The Taylor expansion for a sufficiently arbitrary function $\Phi(x)$
can be expressed as
\[ \Phi(x+\alpha)=\sum_{k=0}^{\infty} \frac{1}{k!}\alpha^k \frac{d^k}{dx^k} \Phi(x)= \exp{\{\alpha \frac{d}{dx}\}}\Phi(x)\ .
\]
Here $\alpha$  can be either a constant or a function of arguments
which do not include $x$ \,(in the second case $\frac{d}{dx}$ is
replaced by $\frac{\partial}{\partial x}$) . If we read this
expression from the right side, we can obtain the result of the
action of an exponential form of a \emph {shift operator}
$\exp{\{\alpha \frac{d}{dx}\}}$  on a function $\Phi(x)$ (see also
(\ref{homom}) and (\ref{homom0}) below):
\[ \exp{\{\alpha \frac{d}{dx}\}}\Phi(x) = \Phi(\exp{\{\alpha \frac{d}{dx}\}}x)= \Phi(x+\alpha) \]
or
\begin{equation} \exp{\{\alpha(\tau) \frac{\partial}{\partial x}\}}\Phi(x) =  \Phi(\exp{\{\alpha(\tau) \frac{\partial}{\partial x}\}}x)=\Phi(x+\alpha(\tau)). \label{Shift1}
\end{equation}

Let us consider shift operators with a change of variable $y=\psi(x)$ . It
is obvious, that
\[ \exp{\{\alpha(\tau) \frac{\partial}{\partial y}\}}\Phi(y) = \exp{\{\alpha(\tau) \frac{\partial}{\partial \psi(x)}\}}\Phi(\psi(x))=\]
\[\exp{\{\alpha(\tau)\frac{1}{\psi(x)'} \frac{\partial}{\partial x}\}}\Phi(\psi(x))=\Phi(\psi(x)+\alpha(\tau)). \]
When $\psi(x)$ is fixed function, the set of shift operators with
different parameters $\alpha(\tau)$  forms an Abelian group. Shift
operators with different $\psi(x)$ do not commute.

Simple examples of shift operators are well-known. We give here
examples which are somewhat exotic
\begin{equation}
\exp{\{\ln(\alpha)x\ln(x) \frac{\partial}{\partial x}\}}\,\,\Phi
(x)= \Phi (x^\alpha)\notag
\end{equation}
and
\begin{equation}
\exp\{\frac{c}{2}\frac{\partial}{\partial
x}\}\,\,\exp{\{\ln(2)x\ln(x) \frac{\partial}{\partial
x}\}}\,\,\exp\{(b-\frac{c^2}{4})\frac{\partial}{\partial
x}\}\,\,\Phi (x)= \Phi (x^2+cx+b)\,.\notag
\end{equation}
These examples demonstrate that combinations of shift operators may
produce non-trivial changes of the variable $x$ in the function $\Phi (x)$.

The following useful identities for shift operators are known
\begin{equation}
\exp{\{a(\tau)x^\alpha \frac{\partial}{\partial x}\}}\,\,x^\beta
\frac{\partial}{\partial x}\,\,\exp{\{-a(\tau)x^\alpha
\frac{\partial}{\partial x}\}}=\notag
\end{equation}
\begin{equation}
\exp{\{a(\tau)x^\alpha \frac{\partial}{\partial
x}\}}\,\,x^{\beta-\alpha }\,\,\exp{\{-a(\tau)x^\alpha
\frac{\partial}{\partial x}\}}\,\,x^\alpha \frac{\partial}{\partial
x}=\notag
\end{equation}
\begin{equation}
\left\{ \begin{array}{rll} \exp\{(\beta-1)a(\tau)\}x^\beta
\frac{\partial}{\partial x}& \mbox{if} & \alpha=1; \\ \\x^\alpha
\frac{\partial}{\partial x}& \mbox{if} & \alpha=\beta;\\ \\
{[x^{\frac{\beta(1-\alpha)}{\beta-\alpha}}+(1-\alpha)a(\tau)x^{\frac{\alpha(1-\alpha)}{\beta-\alpha}}]}^{\frac{\beta-\alpha}{1-\alpha}}\frac{\partial}{\partial
x}& \mbox{if} & \alpha \neq {1}\,\,\mbox{and}\,\, \alpha \neq
\beta.
\end{array}\right.
\label{shiftid}
\end{equation}
Alas, in this last type of transformation the ``polynomial" nature of
the left-hand side of the identities in general does not survive in the
right-hand side.

\subsection{Chronological operator homomorphism}

As we will see later, the chronological operators with the \emph
{derivative} operator ${\bf \Delta}(t)$ (see definition in Section 2)
in the exponential which we will here for short denote as
\begin{equation}
{\bf E} ={\bf T}\exp \{\int_a^t d\tau\,{\bf \Delta}(\tau)\},
\notag
\end{equation}
play a very important role.

Let ${\bf B}(t)=b(t)$ be a \emph{function} that is the operator of
multiplication on the $b(t)$. Then, as can be seen from the definition of
the derivative operator, the commutator $[b(t),{\bf \Delta}(\tau)]$
is a function too.

Now consider the following construction
\begin{equation}
{\bf K}={\bf E}\,b(t)\,{\bf E}^{-1}. \notag
\end{equation}
As far as the commutator $[b(t),{\bf \Delta}(\tau)]$ is a function,
then all repeated commutators are functions too. So with help of BCH
formula (\ref{BprABCH}) we conclude that operator ${\bf K}$ is \emph
{a function} and
\begin{equation}
{\bf K}=({\bf E}\,b(t)). \label{hom}
\end{equation}
An analogous conclusion holds for ${\bf E}^{-1}\,b(t)\,{\bf E}$.

Let us consider further the following obvious chain
\begin{align}
{\bf E}\,b_1\,b_2\,\dots\,b_n&={\bf E}\,b_1\,{\bf E}^{-1}\,{\bf
E}\,b_2\,{\bf E}^{-1}\,\dots\,{\bf E}\,b_n=\notag
\\
&=({\bf E}\,b_1)\,({\bf E}\,b_2)\,\dots\,(\,{\bf E}\,b_n). \notag
\end{align}
With what is observed here and the fact that the chronological operator is linear,
leads to conclusion that for any function
$F(b_1,\,b_2,\,\dots,\,b_n)$, which can be expanded in power series
with respect to $b_1,\,b_2,\,\dots,\,b_n$, we can obtain the
following nice property:
\begin{equation}
{\bf E}\,\,F(b_1,\,b_2,\,\dots,\,b_n)=F(({\bf E}\,b_1),\,({\bf
E}\,b_2),\,\dots,\,({\bf E}\,b_n)). \label{homom}
\end{equation}

Analogously
\begin{equation}
{\bf E}^{-1}\,\,F(b_1,\,b_2,\,\dots,\,b_n)=F(({\bf
E}^{-1}\,b_1),\,({\bf E}^{-1}\,b_2),\,\dots,\,({\bf E}^{-1}\,b_n)).
\label{homom0}
\end{equation}

\subsection{Some additional useful identities}
By taking inverses of both sides of (\ref{BprA}) we find the similar identity for the product of inverse ordered operators
\begin{align}
&{\bf T}_0\exp \{\int_a^t d\tau\,{\bf B}(\tau)\} \,{\bf T}_0\exp
\{\int_a^t d\tau\,{\bf A}(\tau)\}=\notag \\ &{\bf T}_0\exp
\{\int_a^t d\tau\,[{\bf B}(\tau)+{\bf T}\exp \{-\int_a^\tau
d\xi\,{\bf B}(\xi)\}{\bf A}(\tau){\bf T}_0\exp \{\int_a^\tau
d\xi\,{\bf B}(\xi)\}] \}. \notag
\end{align}

Furthermore, the analog of (\ref{BsumA}) is
\begin{align}
{\bf T}_0\exp & \{\int_a^t d\tau\,[{\bf B}(\tau)+{\bf C}(\tau)]\} =
\notag \\ & {\bf T}_0\exp \{\int_a^t d\tau\,{\bf T}_0\exp
\{\int_a^\tau d\xi\,{\bf B}(\xi)\}{\bf C}(\tau){\bf T}\exp
\{-\int_a^\tau d\xi\,{\bf B}(\xi)\} \} \times \notag \\ & {\bf
T}_0\exp \{\int_a^t d\tau\,{\bf B}(\tau)\}. \label{BsumA0}
\end{align}

If in (\ref{bprA}) we substitute
\begin{equation}
{\bf b}(t)={\bf T}_0\exp \{\int_a^t d\tau\,{\bf B}(\tau)\},
\notag
\end{equation}
then from (\ref{bprA}) we obtain the following operator identity
\begin{align}
&{\bf T}_0\exp \{\int_a^t d\tau\,{\bf B}(\tau)\} \,{\bf T}\exp
\{\int_a^t d\tau\,{\bf A}(\tau)\}=\notag \\ &{\bf T}\exp \{\int_a^t
d\tau\,{\bf T}_0\exp \{\int_a^\tau d\xi\,{\bf B}(\xi)\}\,[{\bf
A}(\tau)+{\bf B}(\tau)]{\bf T}\exp \{-\int_a^\tau d\xi\,{\bf
B}(\xi)\}] \}, \notag
\end{align}
and by inverting it and making the change of operators ${\bf A}\rightarrow -{\bf B}$ and ${\bf
B}\rightarrow -{\bf A}$ we obtain another form
\begin{align}
&{\bf T}_0\exp \{\int_a^t d\tau\,{\bf B}(\tau)\} \,{\bf T}\exp
\{\int_a^t d\tau\,{\bf A}(\tau)\}=\notag \\ &{\bf T}_0\exp
\{-\int_a^t d\tau\,{\bf T}_0\exp \{-\int_a^\tau d\xi\,{\bf
A}(\xi)\}\,[{\bf A}(\tau)+{\bf B}(\tau)]{\bf T}\exp \{\int_a^\tau
d\xi\,{\bf A}(\xi)\}] \}. \notag
\end{align}

Since from (\ref{BsumA}) we have that
\begin{align}
{\bf T}\exp & \{\int_a^t d\tau\,[{\bf B}(\tau)+{\bf C}(\tau)]\}
={\bf T}\exp \{\int_a^t d\tau\,{\bf B}(\tau)\} \times \notag \\ &
{\bf T}\exp \{\int_a^t d\tau\,{\bf T}_0\exp \{-\int_a^\tau
d\xi\,{\bf B}(\xi)\}\,{\bf C}(\tau)\,{\bf T}\exp \{\int_a^\tau
d\xi\,{\bf B}(\xi)\} \}=\notag \\ &{\bf T}\exp \{\int_a^t
d\tau\,{\bf C}(\tau)\} \times \notag \\ & {\bf T}\exp \{\int_a^t
d\tau\,{\bf T}_0\exp \{-\int_a^\tau d\xi\,{\bf C}(\xi)\}\,{\bf
B}(\tau)\,{\bf T}\exp \{\int_a^\tau d\xi\,{\bf C}(\xi)\} \}. \notag
\end{align}
then we come to a third form
\begin{align}
&{\bf T}_0\exp \{-\int_a^t d\tau\,{\bf C}(\tau)\}\, {\bf T}\exp
\{\int_a^t d\tau\,{\bf B}(\tau)\}=\notag \\ &{\bf T}\exp \{\int_a^t
d\tau\,{\bf T}_0\exp \{-\int_a^\tau d\xi\,{\bf C}(\xi)\}\,{\bf
B}(\tau)\,{\bf T}\exp \{\int_a^\tau d\xi\,{\bf C}(\xi)\} \}\times
\notag \\ & {\bf T}_0\exp \{-\int_a^t d\tau\,{\bf T}_0\exp
\{-\int_a^\tau d\xi\,{\bf B}(\xi)\}\,{\bf C}(\tau)\,{\bf T}\exp
\{\int_a^\tau d\xi\,{\bf B}(\xi)\} \}. \notag
\end{align}

Let us now consider the invertible operator ${\bf B}$ which \emph {does not depend} on
$t$. Then from (\ref{bprA}) we have
\begin{equation}
{\bf B} \,{\bf T}\exp \{\int_a^t d\tau\,{\bf A}(\tau)\}={\bf T}\exp
\{\int_a^t d\tau\,{\bf B}{\bf A}(\tau){\bf B}^{-1}\} \, {\bf B}.
\notag
\end{equation}
or
\begin{equation}
{\bf T}\exp \{\int_a^t d\tau\,{\bf A}(\tau)\}\,{\bf B} ={\bf B}
\,{\bf T}\exp \{\int_a^t d\tau\,{\bf B}^{-1}{\bf A}(\tau){\bf B}\}.
\label{BprAB}
\end{equation}

From the expression
\begin{equation}
{\bf K}(t)={\bf B}\,{\bf T}\exp \{\int_a^t d\tau\,{\bf A}(\tau)\,{\bf B}\}
\notag
\end{equation}
we obtain by differentiation that
\begin{equation}
\frac{\partial {\bf K}(t)}{\partial t}={\bf B}\,\,{\bf A}(t)\,\,{\bf B}\,\,{\bf T}\exp \{\int_a^t d\tau\,{\bf A}(\tau){\bf B}\}={\bf B}\,\,{\bf A}(t)\,\,{\bf K}(t),
\notag
\end{equation}
\begin{equation}
{\bf K}(t)|_{t=a}={\bf B}.
\notag
\end{equation}
Solving the above operator differential equation we find that for any linear (not necessarily invertible) ``$t$-independent" operator ${\bf B}$
\begin{equation}
{\bf B}\,{\bf T}\exp \{\int_a^t d\tau\,{\bf A}(\tau)\,{\bf B}\}={\bf
T}\exp \{\int_a^t d\tau\,{\bf B}\,{\bf A}(\tau)\}\,{\bf B}. \notag
\end{equation}
Analogously,
\begin{equation}
{\bf B}\,{\bf T}_0\exp \{\int_a^t d\tau\,{\bf A}(\tau)\,{\bf
B}\}={\bf T}_0\exp \{\int_a^t d\tau\,{\bf B}\,{\bf A}(\tau)\}\,{\bf
B}. \notag
\end{equation}

\subsection{Example 1. Linear first-order PDE}

The general linear first-order partial differential equation for
$u(t,\vec{\rho})$
\begin{equation}
\frac{\partial u(t,\vec{\rho})}{\partial
t}=\phi(t,\vec{\rho})+f_0(t,\vec{\rho})u(t,\vec{\rho})+f_1(t,\vec{\rho})\,\frac{\partial
u(t,\vec{\rho})}{\partial x_1}+\dots
+f_n(t,\vec{\rho})\,\frac{\partial u(t,\vec{\rho})}{\partial x_n}\,,
\label{Lpde}
\end{equation}
where we abbreviate $\vec{\rho}=x_1,...,x_n$, with initial condition
\begin{equation}
u(t,\vec{\rho})|_{t=a}=v(\vec{\rho}) \notag
\end{equation}
is fitted outright to be solved by the operator method. Its formal
solution follows from (\ref{baseinsol}):
\begin{align}
u(t,\vec{\rho})&={\bf T}\exp \{ \int_a^t
d\tau\,[f_0(\tau,\vec{\rho})+f_1(\tau,\vec{\rho})\frac{\partial}{\partial
x_1}+\dots+f_n(\tau,\vec{\rho})\frac{\partial}{\partial
x_n}]\} \,v(\vec{\rho})+\notag \\
&\int_a^t d\tau\,{\bf T}\exp \{ \int_\tau^t
d\xi\,[f_0(\xi,\vec{\rho})+f_1(\xi,\vec{\rho})\frac{\partial}{\partial
x_1}+\dots+f_n(\xi,\vec{\rho})\frac{\partial}{\partial x_n}]\}
\,\phi(\tau,\vec{\rho}). \notag
\end{align}

We indicate that the above general solution can be
expressed in terms of some particular solutions of
\emph{homogeneous} first-order PDE:
\begin{equation}
\frac{\partial \tilde{u}(t,\vec{\rho})}{\partial
t}=f_1(t,\vec{\rho})\,\frac{\partial
\tilde{u}(t,\vec{\rho})}{\partial x_1}+\dots
+f_n(t,\vec{\rho})\,\frac{\partial \tilde{u}(t,\vec{\rho})}{\partial
x_n}\,. \label{Lpdesh}
\end{equation}

If we introduce the following notation
\begin{equation}
\zeta_i(t,\vec{\rho})={\bf T}\exp \{ \int_a^t
d\tau\,[f_1(\tau,\vec{\rho})\frac{\partial}{\partial
x_1}+\dots+f_n(\tau,\vec{\rho})\frac{\partial}{\partial x_n}]\}
\,x_i\,,\qquad (i=1,\dots, n) \notag
\end{equation}
we can consider it as the \emph{fundamental} set of
\emph{particular} solutions of (\ref{Lpdesh}) with initial
conditions $\zeta_i(t,\vec{\rho})|_{t=a}=x_i$. Since the operator in
the exponential is the derivative, then taking into account
(\ref{homom}), the solution of homogeneous first-order PDE
(\ref{Lpdesh}) may be expressed via $\zeta_i$ as
\begin{equation}
\tilde{u}(t,\vec{\rho})=v(\zeta_1(t,\vec{\rho}),...,\zeta_n(t,\vec{\rho})).
\notag
\end{equation}
The last expression is the \emph{general} solution of the
homogeneous first-order PDE (\ref{Lpdesh}) if $v(\vec{\rho})$
is an arbitrary function.

If we further denote
\begin{equation}
Z_i(t,\tau,\vec{\rho})={\bf T}\exp \{ \int_\tau^t
d\tau\,[f_1(\tau,\vec{\rho})\frac{\partial}{\partial
x_1}+\dots+f_n(\tau,\vec{\rho})\frac{\partial}{\partial x_n}]\}
\,x_i\,,\qquad (i=1,\dots, n), \notag
\end{equation}
we can find by using (\ref{TT}) that
\begin{equation}
\zeta_i(t,\vec{\rho})={\bf T}\exp \{ \int_\tau^t
d\tau\,[f_1(\tau,\vec{\rho})\frac{\partial}{\partial
x_1}+\dots+f_n(\tau,\vec{\rho})\frac{\partial}{\partial x_n}]\}
\,\zeta_i(\tau,\vec{\rho})\,, \notag
\end{equation}
or
\begin{equation}
\zeta_i(t,\vec{\rho})=\zeta_i(\tau,Z_1(t,\tau,\vec{\rho}),\dots,Z_n(t,\tau,\vec{\rho}))\,,
\notag
\end{equation}
so $Z_i(t,\tau,\vec{\rho})$ are solutions of the following system of
algebraic equations
\begin{equation}
\zeta_i(\tau,Z_1,\dots,Z_n)=\zeta_i(t,\vec{\rho}),\qquad (i=1,\dots,
n). \label{zZ}
\end{equation}

Let us take up in passing a nice property of considered type of
functions which we will need later. Denoting
\begin{equation}
b_i(t,\vec{\rho})={\bf T}_0\exp \{ -\int_a^t
d\tau\,[f_1(\tau,\vec{\rho})\frac{\partial}{\partial
x_1}+\dots+f_n(\tau,\vec{\rho})\frac{\partial}{\partial x_n}]\}
\,x_i\,,\qquad (i=1,\dots,n) \label{bi}
\end{equation}
then from (\ref{TT0}) and (\ref{homom0}) we have
\begin{align}
{\bf T}_0\exp \{ -\int_a^t
d\tau\,[f_1(\tau,\vec{\rho})\frac{\partial}{\partial
x_1}&+\dots+f_n(\tau,\vec{\rho})\frac{\partial}{\partial x_n}]\}
\,\zeta_i(t,\vec{\rho})=\notag \\
&\zeta_i(t,b_1(t,\vec{\rho}),\dots,b_n(t,\vec{\rho}))=x_i
\label{alg1}
\end{align}
and analogously
\begin{equation}
b_i(t,\zeta_1(t,\vec{\rho}),\dots,\zeta_n(t,\vec{\rho}))=x_i\,,\qquad
(i=1,\dots,n).\label{alg2}
\end{equation}
That is $\zeta_i$ and $b_i$ are bundled by \emph{algebraic} systems
(\ref{alg1}) and (\ref{alg2}).

Moreover, it is obvious that
\begin{equation}
\frac{\partial b_i(t,\vec{\rho})}{\partial t}=-{\bf T}_0\exp \{
-\int_a^t d\tau\,[f_1(\tau,\vec{\rho})\frac{\partial}{\partial
x_1}+\dots+f_n(\tau,\vec{\rho})\frac{\partial}{\partial x_n}]\}
\,f_i(\tau,\vec{\rho})\,, \notag
\end{equation}
and from (\ref{homom0}) it follows that
\begin{equation}
\frac{\partial b_i(t,\vec{\rho})}{\partial
t}=-\,f_i(t,b_1(t,\vec{\rho}), \dots,b_n(t,\vec{\rho}))\,,\qquad
(i=1,\dots,n)\,. \label{bi2}
\end{equation}
The expressions (\ref{bi}) are solutions of the system of
$n$ first-order \emph{non-linear} ODEs (\ref{bi2}).

As by virtue of (\ref{BsumA}) and (\ref{homom})
\begin{align}
{\bf T}\exp &\{ \int_a^t
d\tau\,[f_0(\tau,\vec{\rho})+f_1(\tau,\vec{\rho})\frac{\partial}{\partial
x_1}+\dots+f_n(\tau,\vec{\rho})\frac{\partial}{\partial x_n}]\}
=\notag \\
&{\bf T}\exp \{ \int_a^t
d\tau\,[f_1(\tau,\vec{\rho})\frac{\partial}{\partial
x_1}+\dots+f_n(\tau,\vec{\rho})\frac{\partial}{\partial
x_n}]\}\times \notag \\
& \exp\{\int_a^t d\tau\,({\bf T}_0\exp \{ \int_a^\tau
d\xi\,[f_1(\xi,\vec{\rho})\frac{\partial}{\partial
x_1}+\dots+f_n(\xi,\vec{\rho})\frac{\partial}{\partial
x_n}]\}f_0(\tau,\vec{\rho}) )\}= \notag
\end{align}
\begin{align}
&\exp\{\int_a^t
d\tau\,f_0(\tau,Z_1(t,\tau,\vec{\rho}),\dots,Z_n(t,\tau,\vec{\rho})
)\} \times \notag\\&{\bf T}\exp \{ \int_a^t
d\tau\,[f_1(\tau,\vec{\rho})\frac{\partial}{\partial
x_1}+\dots+f_n(\tau,\vec{\rho})\frac{\partial}{\partial
x_n}]\}\,.\notag
\end{align}
If we now return to the general linear first-order PDE (\ref{Lpde}), we
can rewrite its operator solution via $\zeta_i$ (and $Z_i$ which are
expressed through $\zeta_i$ by system (\ref{zZ})) into the following
form
\begin{align}
&u(t,\vec{\rho})=v(\zeta_1(t,\vec{\rho}),...,\zeta_n(t,\vec{\rho}))\exp\{\int_a^t
d\tau\,f_0(\tau,Z_1(t,\tau,\vec{\rho}),\dots,Z_n(t,\tau,\vec{\rho})
)\}+\notag \\
&\int_a^t
d\tau\,\phi(\tau,Z_1(t,\tau,\vec{\rho}),\dots,Z_n(t,\tau,\vec{\rho}))\exp\{\int_\tau^t
d\xi\,f_0(\xi,Z_1(t,\xi,\vec{\rho}),\dots,Z_n(t,\xi,\vec{\rho}) )\}.
\notag
\end{align}

\subsection{Example 2. Linear parabolic differential equation}

The linear parabolic differential equation
\begin{equation}
\frac{\partial u(t,\vec{\rho})}{\partial
t}=f_0(t,\vec{\rho})\,u(t,\vec{\rho})+f_1(t,\vec{\rho})\,\frac{\partial^2
u(t,\vec{\rho})}{\partial x_1^2}+\dots
+f_n(t,\vec{\rho})\,\frac{\partial^2 u(t,\vec{\rho})}{\partial
x_n^2}\,, \notag
\end{equation}
where $\vec{\rho}=x_1,...,x_n$, with an initial condition
\begin{equation}
u(t,\vec{\rho})|_{t=0}=v(\vec{\rho}), \notag
\end{equation}
has the following formal solution
\begin{equation}
u(t,\vec{\rho})={\bf T}\exp \{ \int_0^t
d\tau\,[f_0(\tau,\vec{\rho})+f_1(\tau,\vec{\rho})\frac{\partial^2}{\partial
x_1^2}+\dots+f_n(\tau,\vec{\rho})\frac{\partial^2}{\partial
x_n^2}]\} \,v(\vec{\rho}). \notag
\end{equation}

It is interesting to note that it is generally accepted that
the parabolic equation is a \emph{second} order PDE, but from the angle of
the operator method considered here this equation is \emph{first}
order \emph{if} the initial condition is specified for $t$ variable.

In the simplest case when all $f_i$ are constants $f_i = k_i$ then it
is easy to see with (\ref{BprAB}) that
\begin{align}
&{\bf T}\exp \{ \int_0^t d\tau\,[k_0+k_1\frac{\partial^2}{\partial
x_1^2}+\dots+k_n\frac{\partial^2}{\partial x_n^2}]\}
\,\exp\{i(\sigma_1 x_1+\dots+\sigma_n x_n)\}=\notag \\
&\exp\{i(\sigma_1 x_1+\dots+\sigma_n x_n)\}\, \exp \{
t\,[k_0-k_1\sigma_1^2-\dots-k_n\sigma_n^2]\}\,,\notag
\end{align}
so we can obtain well-known non-operator solutions of such a
parabolic equation with help of a Fourier
transformation of the initial condition.

\section{Non-linear first-order ODE's}
\subsection{The operator solution for first-order non-linear ODE}

An ordinary \emph{non-linear} first-order differential equation
\begin{equation}
\frac{du(t)}{dt}=f(t,u(t)) \label{ode1}
\end{equation}
on account of its non-linearity is not immediately suited for
application of the above considered operator method. Nevertheless there
are many possibilities to convert the problem (\ref{ode1}) to a linear
one and in Subsection 3.9 we \emph{have obtained} by way of the
operator method a solution for such an equation. Here we consider this
problem in more detail.

Most linearization procedures  are concerned with the introduction of
spaces of larger dimensions. Here we demonstrate some of them which
are almost generally applicable.

If we introduce a new function
\begin{equation}
S(t,\omega)=e^{\omega u(t)}\,, \label{S}
\end{equation}
from (\ref{ode1}) one can
easily derive the following first-order (with respect to $t$)
linear differential equation for $S(t,\omega)$ equivalent to (\ref{ode1})
\begin{equation}
\frac{\partial S(t,\omega)}{\partial t}=\omega
f(t,\frac{\partial}{\partial \omega})S(t,\omega)\ , \qquad
S(a,\omega)=e^{\omega u(a)}, \label{Lode1}
\end{equation}
from which the formal solution follows immediately from (\ref{uT}) in the
form
\begin{equation}
S(t,\omega)={\bf T}\, \exp\{\int_a^t d\tau\,\omega
f(\tau,\frac{\partial }{\partial \omega})\}e^{\omega u(a)}.
\label{Lode1sol}
\end{equation}
So the solution of equation (\ref{ode1}) is:
\begin{equation}
u(t)=\frac{\partial S(t,\omega)}{\partial \omega}|_{\omega=0}.
\label{Nsol}
\end{equation}

Direct substitution of the solution (\ref{Nsol}) reduces the
equation (\ref{ode1}) to an identity.

The theorem for the uniqueness of a solution of a differential equation
shows that if we have an exact solution of a problem in several
forms, all the forms can be transformed to each other. So the main
problem that remains in this approach is how to transform the
operator solution to that form which can be considered most useful.

\subsection{Solution for first-order ODE in form with derivative
operator}

The main result of this subsection is another form (\ref{Lode2sol}) of
the formal solution for non-linear first-order ODE. We can prove the
solution correctness by its direct substitution into ODE
(\ref{ode1}), but it is desirable to outline the way which leads to
such form.

Let us start from an operator solution of the equation (\ref{Lode1}) in
the form (\ref{Lode1sol}) but let us rewrite it (not only) for
convenience in the following notation
\[S(t,c,\omega)={\bf T}\, \exp\{\int_a^t d\tau\,\omega f(\tau,\frac{\partial }{\partial \omega})\}\,e^{\omega c},\]
where $c=u(t)|_{t=a}$. With the help of identity (\ref{BprAB}) we find
\[S(t,c,\omega)=e^{\omega c}\,{\bf T}\, \exp\{\int_a^t d\tau\,\omega \,e^{-\omega c} f(\tau,\frac{\partial }{\partial \omega})e^{\omega c}\}\cdot1.\]
CHB-expansion gives
\begin{equation}
e^{-\omega c} f(\tau,\frac{\partial }{\partial \omega})e^{\omega
c}=f(\tau,\frac{\partial }{\partial \omega})+c[f(\tau,\frac{\partial
}{\partial \omega}),\omega]+\frac{c^2}{2}[[f(\tau,\frac{\partial
}{\partial \omega}),\omega],\omega]+\dots \label{Expr}
\end{equation}
Since it is easily proven by induction that
\[[\frac{\partial^m }{\partial \omega^m},\omega]=m\frac{\partial^{m-1} }{\partial \omega^{m-1}},\]
then from expansion of $f(x)$ into power series and backwards
summation it follows that
\[[f(\frac{\partial }{\partial \omega}),\omega]=f'(\frac{\partial }{\partial \omega})\qquad (f'(x)=\frac{\partial f(x) }{\partial x}),\]
therefore we can conclude with taking into account (\ref{Expr}) and
the property of the shift operator (\ref{Shift1}) that
\[e^{-\omega c} f(\tau,\frac{\partial }{\partial \omega})e^{\omega c}=f(\tau,\frac{\partial }{\partial \omega})+c\,f'(\tau,\frac{\partial }{\partial \omega})+\frac{c^2}{2}\,f''(\tau,\frac{\partial }{\partial \omega})+\dots=\]

\[f(\tau,c+\frac{\partial }{\partial \omega})=\exp\{\frac{\partial }{\partial \omega}\frac{\partial }{\partial c}\} f(\tau,c)  \exp\{-\frac{\partial }{\partial \omega}\frac{\partial }{\partial c}\}.\]
Then
\begin{align}
S(t,c,\omega)& =e^{\omega c}\exp\{\frac{\partial }{\partial
\omega}\frac{\partial }{\partial c}\}{\bf T}\, \exp\{\int_a^t
d\tau\,(\omega-\frac{\partial }{\partial c})
f(\tau,c)\}\,\exp\{-\frac{\partial }{\partial \omega}\frac{\partial
}{\partial c}\}\cdot 1 =\notag \\ &  e^{\omega
c}\exp\{\frac{\partial }{\partial \omega}\frac{\partial }{\partial
c}\}{\bf T}\, \exp\{\int_a^t d\tau\,[\omega f(\tau,c)-\frac{\partial
f(\tau,c)}{\partial c} -f(\tau,c)\frac{\partial}{\partial c}] \}
\cdot 1. \notag
\end{align}
By expanding the chronological exponential of the
derivative operator $f(\tau,c)\frac{\partial}{\partial c}$ with the help of identity
(\ref{BsumA}) and properties (\ref{hom}), (\ref{homom})\,  we obtain
that
\begin{align}
S( t,c,\omega)& =e^{\omega c}\exp\{\frac{\partial }{\partial
\omega}\frac{\partial }{\partial c}\}{\bf T}\, \exp\{-\int_a^t
d\tau\,f(\tau,c)\frac{\partial}{\partial c}\} \times \notag \\ &
 \exp\{-\int_a^t d\tau\,g(\tau,c)\} \exp\{\omega\int_a^t
d\tau\,G(\tau,c)\},\notag
\end{align}
where
\[ g(\tau,c)={\bf T}_0\, \exp\{\int_a^\tau d\xi\,f(\xi,c)\frac{\partial}{\partial c}\}\,\frac{\partial f(\tau,c)}{\partial c} \]
and
\[ G(\tau,c)={\bf T}_0\, \exp\{\int_a^\tau d\xi\,f(\xi,c)\frac{\partial}{\partial c}\}\,f(\tau,c). \]
If we differentiate the last expression with respect to $\omega$ we
find that
\begin{align}
\frac{\partial S(t,c,\omega) }{\partial
\omega}&=c\,S(t,c,\omega)+e^{\omega c}\exp\{\frac{\partial
}{\partial \omega}\frac{\partial }{\partial c}\}{\bf T}\,
\exp\{-\int_a^t d\tau\,f(\tau,c)\frac{\partial}{\partial
c}\}\times\notag \\ & \exp\{-\int_a^t d\tau\,g(\tau,c)\}
\exp\{\omega\int_a^t d\tau\,G(\tau,c)\}\,\int_a^t d\tau\,G(\tau,c).
\notag
\end{align}
If in the second item we recover the initial operator form
\[\frac{\partial S(t,c,\omega) }{\partial \omega}=c\,S(t,c,\omega)+ {\bf T}\, \exp\{\int_a^t d\tau\,\omega f(\tau,\frac{\partial }{\partial \omega})\}\,e^{\omega c}\,\int_a^t d\tau\,G(\tau,c), \]
and note that the chronological operator commutes with any function
which does not depend on $\omega$, we arrive at
\[\frac{\partial S(t,c,\omega) }{\partial \omega}={[c+ \int_a^t d\tau\,G(\tau,c)}]\,S(t,c,\omega) \]
and inasmuch as $S(t,c,\omega)|_{\omega=0}\equiv 1$, then
\[ u(t,c)= \frac{\partial S(t,c,\omega) }{\partial \omega}|_{\omega=0}=c+ \int_a^t d\tau\,G(\tau,c),\]
hence it follows that
\[ u(t,c)=c+ \int_a^t d\tau\,{\bf T}_0\, \exp\{\int_a^\tau d\xi\,f(\xi,c)\frac{\partial}{\partial c}\}\,f(\tau,c)\]
or finally
\begin{equation}
u(t,c)={\bf T}_0\, \exp\{\int_a^t
d\tau\,f(\tau,c)\,\frac{\partial}{\partial c}\}\,c. \label{Lode2sol}
\end{equation}
The expression (\ref{Lode2sol}) is the \emph{general} solution of
the equation (\ref{ode1}) if we will consider $c$ as an arbitrary
constant.

Above derivation is not rigorous, of course, so it is very
important to verify our conclusion by direct substitution of
the obtained solution into equation (\ref{ode1}). After differentiation
of (\ref{Lode2sol}) and using property (\ref{homom}) we can be sure
that (\ref{Lode2sol}) is really the formal solution of the equation
(\ref{ode1}).

So we have obtained a solution for a non-linear first-order equation in
a more convenient form with the derivative operator.

Let us finish this Subsection with a brief remark about the well-known
interconnection between first-order ODEs and linear first-order PDEs.

As we have seen in Example 1 above (see Subsection 3.9) the function
\begin{equation}
\zeta(t,c)={\bf T}\, \exp\{-\int_a^t
d\tau\,f(\tau,c)\frac{\partial}{\partial c}\}\,c \label{Pdesol}
\end{equation}
satisfies the equation
\[\frac{\partial z(t,c)}{\partial t}+f(t,c)\,\frac{\partial z(t,c)}{\partial c}=0.\]
If we now act on both sides of (\ref{Pdesol}) by the operator
\[ {\bf T}_0\, \exp\{\int_a^t d\tau\,f(\tau,c)\frac{\partial}{\partial c}\}, \]
we get
\begin{equation}
\zeta(t,u(t,c))=c  \label{Pdez}
\end{equation}
and
\[ u(t,\zeta(t,c))=c  \]
 -- the well-known equations expressing the relationship between
solutions of first-order ODEs and the corresponding linear first-order
PDEs.

In some practical cases it is easier to obtain explicit expression
of $\zeta(t,c)$ for given ODE, for example, by an integrating factor
method. (\ref{Pdez}) then is the implicit form of the solution for
the given ODE. Which form of solution of types (\ref{Lode2sol}) or
(\ref{Pdez}) is more analyzable depends on the object under investigation for a
specific problem.

\subsection{Example 3. Bernoulli ODE}

The formal solution of a Bernoulli ODE
\begin{equation}
\frac{du}{dt}=a(t)u^\alpha+b(t)u \notag
\end{equation}
in the form of (\ref{Lode2sol}) is as follows
\begin{equation}
u(t,c)={\bf T}_0\, \exp\{\int_0^t
d\tau\,[a(\tau)c^\alpha+b(\tau)c]\,\frac{\partial}{\partial c}\}\,c.
\notag
\end{equation}
Expanding the above chronological exponential with the help of identity
(\ref{BsumA0})
\begin{align}
u(t,c)={\bf T}_0\, \exp\{\int_0^t d\tau\,a(\tau)\,
&\exp\{\int_0^\tau d\xi\,b(\xi)c\,\frac{\partial}{\partial
c}\}\,c^\alpha\,\frac{\partial}{\partial c}\,\exp\{-\int_0^\tau
d\xi\,b(\xi)c\,\frac{\partial}{\partial c}\}\}\,\times \notag \\
& \exp\{\int_0^t d\tau\,b(\tau)c\,\frac{\partial}{\partial c}\}\,c
\notag
\end{align}
and with (\ref{shiftid}) gives us the solution with only shift
operators
\begin{equation}
u(t,c)=\exp\{\int_0^t d\tau\,a(\tau)\, \exp\{(\alpha -1)\int_0^\tau
d\xi\,b(\xi)\}\,c^\alpha\,\frac{\partial}{\partial c}\}\,
\exp\{\int_0^t d\tau\,b(\tau)c\,\frac{\partial}{\partial c}\}\,c\,.
\notag
\end{equation}
Executing shift operations we arrive to a classical form of the solution
for Bernoulli ODE ($\alpha\neq1$)
\begin{equation}
u(t,c)=[c^{(1-\alpha)}+(1-\alpha)\int_0^t d\tau\,a(\tau)\,
\exp\{(\alpha -1)\int_0^\tau d\xi\,b(\xi)\}]^{\frac{1}{
1-\alpha}}\exp\{\int_0^t d\tau\,b(\tau)\}. \notag
\end{equation}

The technique being used here is applicable not only for Bernoulli
ODEs, but it is successful under concatenation of some circumstances,
when the original chronological exponential is decomposed into the chain of
shift operators on its own without solving any auxiliary
differential equations.

\subsection{The ``integral-free'' form of solution for non-linear first-order ODE}

Let us consider the following operator chain
\begin{align}
&\exp\{(t-a)\,[f(s,c)\,\frac{\partial}{\partial
c}+\frac{\partial}{\partial s}]\}\overset{(\ref{BsumA0})}{=}\notag \\
& {\bf T}_0\, \exp\{\int_a^t
d\tau\,e^{\{(\tau-a)\frac{\partial}{\partial
s}\}}f(s,c)\,\frac{\partial}{\partial
c}e^{\{(a-\tau)\frac{\partial}{\partial
s}\}}\}\,\exp\{(t-a)\frac{\partial}{\partial
s}\}\overset{(\ref{Shift1})}{=}\notag
\\ & {\bf
T}_0\, \exp\{\int_a^t d\tau\,f(s+\tau-a,c)\,\frac{\partial}{\partial
c}\}\,\exp\{(t-a)\frac{\partial}{\partial s}\}. \notag
\end{align}

So
\begin{equation}
{\bf T}_0\, \exp\{\int_a^t
d\tau\,f(s+\tau-a,c)\,\frac{\partial}{\partial c}\}\,c
=\exp\{(t-a)\,[f(s,c)\,\frac{\partial}{\partial
c}+\frac{\partial}{\partial s}]\}\, c\notag
\end{equation}
and if we denote
\begin{equation}
U(t,c,s) =\exp\{(t-a)\,[f(s,c)\,\frac{\partial}{\partial
c}+\frac{\partial}{\partial s}]\}\, c\notag
\end{equation}
we obtain from the preceding expression that the general solution of ODE
(\ref{ode1}) is
\begin{equation}
u(t,c) =U(t,c,s)|_{s=a}\notag
\end{equation}
or
\begin{equation}
u(t,c) =[\exp\{(t-a)\,[f(s,c)\,\frac{\partial}{\partial
c}+\frac{\partial}{\partial s}]\}\, c]|_{s=a}\,.\notag
\end{equation}

The last form of the operator solution has some interesting features in
the sense that it does not contain any integration and even ordering
operator ${\bf T}$, which may be useful for calculating
\emph{approximate} expressions of the solution $u(t,c)$.

The function $U(t,c,s)$ obviously satisfies the following PDE
\begin{equation}
\frac{\partial U(t,c,s)}{\partial t}-f(s,c)\,\frac{\partial
U(t,c,s)}{\partial c}-\frac{\partial U(t,c,s)}{\partial
s}=0\,,\qquad (U(t,c,s)|_{t=a}=c). \notag
\end{equation}

\subsection{The solution with an arbitrary function}

Let us now consider $f(t,c)=g(t,c)+h(t,c)$ so we can find the following form
of the general solution of equation (\ref{ode1})
\begin{align}
u(t,c)&={\bf T}_0\, \exp\{\int_a^t
d\tau\,f(\tau,c)\,\frac{\partial}{\partial c}\}\,c\,={\bf T}_0\,
\exp\{\int_a^t
d\tau\,[g(\tau,c)+h(\tau,c)]\,\frac{\partial}{\partial
c}\}\,c=\notag \\ & {\bf T}_0\, \exp\{\int_a^t d\tau\, [{\bf T}_0\,
\exp\{\int_a^\tau d\xi\,h(\xi,c)\,\frac{\partial}{\partial
c}\}\,g(\tau,c)]\,\times\notag \\ & \exp\{-\int_a^\tau d\xi \, [{\bf
T}_0\,\exp\{\int_a^\xi d\zeta \,h(\zeta,c)\,\frac{\partial}{\partial
c}\}\,\frac{\partial h(\xi,c)}{\partial
c}]\}\frac{\partial}{\partial c}\}\,\times\notag
\\ &{\bf T}_0\, \exp\{\int_a^t
d\tau\,h(\tau,c)\,\frac{\partial}{\partial c}\}\,c \notag
\end{align}
or
\begin{align}
u(t,c)&={\bf T}_0\, \exp\{\int_a^t d\tau\, [{\bf T}_0\,
\exp\{\int_a^\tau d\xi\,h(\xi,c)\,\frac{\partial}{\partial
c}\}\,[f(\tau,c)-h(\tau,c)]]\,\times\notag \\ & \exp\{-\int_a^\tau
d\xi \, [{\bf T}_0\, \exp\{\int_a^\xi d\zeta
\,h(\zeta,c)\,\frac{\partial}{\partial c}\}\,\frac {\partial
h(\xi,c)}{\partial c}]\}\frac{\partial}{\partial c}\}\,\times\notag
\\ &{\bf T}_0\, \exp\{\int_a^t
d\tau\,h(\tau,c)\,\frac{\partial}{\partial c}\}\,c, \notag
\end{align}
where we can consider $h(t,c)$ as an arbitrary differentiable
function.

Supposing it is known that
\[ z(t,c)={\bf T}_0\, \exp\{\int_a^t d\tau\,h(\tau,c)\,\frac{\partial}{\partial c}\}\,c, \]
then with the help of property (\ref{homom}) we obtain
\begin{equation}
u(t,c)={\bf T}_0\, \exp\{\int_a^t
d\tau\,\frac{f(\tau,z(\tau,c))-\frac{\partial z(\tau,c)}{\partial
\tau}}{\frac{\partial z(\tau,c)}{\partial
c}}\,\frac{\partial}{\partial c}\}\,z(t,c). \label{Asol2}
\end{equation}
As far as $h(t,c)$ is an arbitrary function, the preceding
expression is valid for any differentiable function $z(t,c)$. The
particular solution of equation (\ref{ode1}) with initial condition
$u(t,c)|_{t=a}=c$ is obtained when $z(t,c)|_{t=a}=c$.

Rewriting (\ref{Asol2}) as
\begin{equation}
u(t,c)=z(t,{\bf T}_0\, \exp\{\int_a^t
d\tau\,\frac{f(\tau,z(\tau,c))-\frac{\partial z(\tau,c)}{\partial
\tau}}{\frac{\partial z(\tau,c)}{\partial
c}}\,\frac{\partial}{\partial c}\}\,c) \label{Asol3}
\end{equation}
by wisely choosing the form of the function $z(t,c)$ we may reduce the initial
problem to
\begin{equation}
\tilde{u}(t,c)={\bf T}_0\, \exp\{\int_a^t
d\tau\,\frac{f(\tau,z(\tau,c))-\frac{\partial z(\tau,c)}{\partial
\tau}}{\frac{\partial z(\tau,c)}{\partial
c}}\,\frac{\partial}{\partial c}\}\,c, \notag
\end{equation}
which may be a simpler one by losing some troublesome singularities of
the initial problem.

\section{The systems of non-linear first order ODEs}

\subsection{The formal solution of the system of non-linear first order ODEs}

To find the formal solution of the \emph{system} of non-linear first
order ODEs
\begin{equation}
\frac{du_i(t)}{dt}=f_i(t,u_1(t),\dots,u_n(t)),\qquad (i=1,\dots,n)
\label{sys}
\end{equation}
we, of course, could introduce an auxiliary function like (\ref{S})
and pass through a chain of unwieldy expressions but now we are ready
to assert that the general solution of the system (\ref{sys}) has
the following operator form
\begin{equation}
u_i(t,\vec{c})={\bf T}_0\exp \{ \int_a^t
d\tau\,[f_1(\tau,\vec{c})\frac{\partial}{\partial
c_1}+\dots+f_n(\tau,\vec{c})\frac{\partial}{\partial c_n}]\}
\,c_i\,, \label{syssol}
\end{equation}
where we abbreviate $\vec{c}=c_1,...,c_n$ and $c_i$ are a set of
arbitrary constants.

Differentiating (\ref{syssol}) with respect to $t$ and with
(\ref{homom0}) we have
\begin{align}
\frac{\partial u_i}{\partial t}&={\bf T}_0\exp \{ \int_a^t
d\tau\,[f_1(\tau,\vec{c})\frac{\partial}{\partial
c_1}+\dots+f_n(\tau,\vec{c})\frac{\partial}{\partial c_n}]\}
\,f_i(t,\vec{c})=\notag\\&= f_i(t,u_1,\dots,u_n) . \notag
\end{align}

If we denote here that
\begin{equation}
\zeta_\phi(t,\vec{c})={\bf T}\exp \{-\int_a^t
d\tau\,[f_1(\tau,\vec{c})\frac{\partial}{\partial
c_1}+\dots+f_n(\tau,\vec{c})\frac{\partial}{\partial c_n}]\}
\,\phi(\vec{c})\,, \notag
\end{equation}
it is easy to see that $\zeta_\phi$ for any function $\phi$ satisfies
a PDE similar to (\ref{Lpdesh}):
\begin{equation}
\frac{\partial \zeta_\phi}{\partial t}+f_1(t,\vec{c})\frac{\partial
\zeta_\phi}{\partial c_1}+\dots+f_n(t,\vec{c})\frac{\partial
\zeta_\phi}{\partial c_n} \,=\,0\,. \notag
\end{equation}
So if we know the $n$ fundamental solutions of this equation, namely
\begin{equation}
\zeta_i(t,\vec{c})={\bf T}\exp \{-\int_a^t
d\tau\,[f_1(\tau,\vec{c})\frac{\partial}{\partial
c_1}+\dots+f_n(\tau,\vec{c})\frac{\partial}{\partial c_n}]\}
\,c_i\,, \notag
\end{equation}
then
\[\zeta_\phi(t,\vec{c})=\phi(\zeta_1(t,\vec{c}), \dots,\zeta_n(t,\vec{c}))\]
and we can find $u_i$ as the solution of the algebraic system
\begin{equation}
\zeta_i(t,u_1,\dots,u_n)=c_i\,,\qquad (i=1,\dots,n).\notag
\end{equation}

Analogous to the first order ODE, we can rewrite the solution
(\ref{syssol}) in an ``integral-free" form as follows
\begin{equation}
u_i(t,\vec{c})=[\exp \{
(t-a)\,[f_1(s,\vec{c})\frac{\partial}{\partial
c_1}+\dots+f_n(s,\vec{c})\frac{\partial}{\partial
c_n}+\frac{\partial}{\partial s}]\} \,c_i\,]|_{s=a}\,. \notag
\end{equation}

\subsection{The direct calculation of BCH type expressions when the involved operators are derivatives}

As we have seen, the operators similar to (\ref{BCH0}) when operators ${\bf
A}(t)$ and ${\bf B}(t)$ are derivatives similar to (\ref{L1}), play
a key role in our approach. And one of the principal points here is
the calculation of the following expressions
\begin{align}
&{\bf K}_i(t,\vec{x})=\notag\\&{\bf T}_0\exp \{- \int_a^t
d\tau\,\sum_{j=1}^m h_j(\tau,\vec{x})\frac{\partial}{\partial x_j}\}
\,\frac{\partial}{\partial x_i}\,{\bf T}\exp \{\int_a^t
d\tau\,\sum_{j=1}^m h_j(\tau,\vec{x})\frac{\partial}{\partial x_j}\}
\,, \label{Kl}
\end{align}
where $\vec{x}=x_1,\dots ,x_m$, $h_j(t,\vec{x})$ are ordinary
functions, and we suppose that for a given $h_j(t,\vec{x})$ we are
able to calculate the fundamental set of of functions
\begin{equation}
z_i(t,\vec{x})={\bf T}_0\exp \{- \int_a^t d\tau\,\sum_{j=1}^m
h_j(\tau,\vec{x})\frac{\partial}{\partial x_j}\}\,x_i\,. \notag
\end{equation}
We have calculated above such expressions for $m=1$, but if
$m>1$ we need to follow by a different way.

Analogous to (\ref{BCH0}), we find that
\begin{align}
&\frac{\partial {\bf K}_i(t,\vec{x})}{\partial t}=\notag\\&{\bf
T}_0\exp \{- \int_a^t d\tau\,\sum_{j=1}^m
h_j(\tau,\vec{x})\frac{\partial}{\partial x_j}\}
\,(\sum_{j=1}^m\frac{\partial h_j(t,\vec{x})}{\partial
x_i}\,\frac{\partial}{\partial x_j})\,\times \notag \\&\,{\bf T}\exp
\{\int_a^t d\tau\,\sum_{j=1}^m
h_j(\tau,\vec{x})\frac{\partial}{\partial x_j}\} \,. \notag
\end{align}
The use of (\ref{homom0}) leads to the following system of
\emph{linear} operator ODEs
\begin{equation}
\frac{\partial {\bf K}_i(t,\vec{x})}{\partial t}=\sum_{j=1}^m
g_{ij}(t,\vec{x}) {\bf K}_j(t,\vec{x})\, \,, \qquad (i=1, \dots,
m)\,, \label{K3}
\end{equation}
where
\begin{equation}
g_{ij}(t,\vec{x})={\bf T}_0\exp \{- \int_a^t d\tau\,\sum_{j=1}^m
h_j(\tau,\vec{x})\frac{\partial}{\partial x_j}\}\,\frac{\partial
h_j(t,\vec{x})}{\partial x_i}\,, \notag
\end{equation}
with initial conditions
\begin{equation}
{\bf K}_i(t,\vec{x})|_{t=a}= \frac{\partial}{\partial x_i}\,.
\label{K4}
\end{equation}

We can write down the formal solution of the system
(\ref{K3})-(\ref{K4}) as
\begin{equation}
{\bf K}_i(t,\vec{x})=[{\bf T}_0\exp \{ \int_a^t
d\tau\,\sum_{l=1}^m\sum_{j=1}^m
g_{lj}(\tau,\vec{x})\,c_j\,\frac{\partial}{\partial
c_l}\}\,c_i]\,|\,_{c_i=\frac{\partial }{\partial x_i}}\,. \notag
\end{equation}
Since the system (\ref{K3}) is linear, then its solutions depend on
initial conditions \emph{linearly}, i.e.
\begin{equation}
{\bf K}_i(t,\vec{x})=\sum_{k=1}^m p_{ik}(t,\vec{x})\,\frac{\partial
}{\partial x_k}\,,  \notag
\end{equation}
where
\begin{equation}
p_{ik}(t,\vec{x})=[{\bf T}_0\exp \{\int_a^t
d\tau\,\sum_{l=1}^m\sum_{j=1}^m
g_{lj}(\tau,\vec{x})\,c_j\,\frac{\partial}{\partial
c_l}\}\,c_i]\,|\,_{c_k=1,\, c_{i\neq k} =0}\,. \notag
\end{equation}
This conclusion is a direct consequence of the fact that the set of
operators $\sum_{j=1}^m h_j(t,\vec{x})\frac{\partial}{\partial x_j}$
forms a Lie algebra.

So in cases under consideration we can exactly calculate the
operators ${\bf K}_i(t,\vec{x})$ (\ref{Kl}) without resorting to BCH
expansions.

The usage of such calculations \emph{in principle} allows us to
factor the chronological exponential, e.g. in (\ref{syssol}), into a product of
relatively simple factors (with a sequence order assigned in
advance)
\begin{equation}
{\bf T}_0\exp \{\int_a^t d\tau\,\sum_{j=1}^n
f_j(\tau,\vec{c})\frac{\partial}{\partial c_j}\}=\prod_{i=1}^n{\bf
T}_0\exp \{\int_a^t d\tau\,
g_i(\tau,\vec{c})\frac{\partial}{\partial c_i}\}\,. \notag
\end{equation}
Unfortunately this way leads to complicated non-linear PDEs for
functions $g_i(t,\vec{c})$.

\section{The non-linear $n$th order ODE}

For the non-linear \emph{$n$th order} ODE
\begin{equation}
\frac{d^nu(t)}{dt^n}=f(t,u(t),\frac{du(t)}{dt},\dots,\frac{d^{n-1}u(t)}{dt^{n-1}})
\label{odeN}
\end{equation}
we can apply the results of the previous Section as long as the $n$th
order ODE can be represented by a system of $n$ first order ODEs:
\begin{align}
&\frac{du(t)}{dt}=u_1(t);\notag\\&........\notag\\&\frac{du_i(t)}{dt}=u_{i+1}(t);\notag\\&........\notag\\&\frac{du_{n-1}(t)}{dt}=f(t,u(t),u_1(t),\dots,u_{n-1}(t))
. \notag
\end{align}
So the formal solution of ODE (\ref{odeN}) can be expressed in the
following form (here $\vec{c}=c_1,...,c_n$)
\begin{equation}
u(t,\vec{c})={\bf T}_0\exp \{ \int_a^t
d\tau\,[f(\tau,\vec{c})\frac{\partial}{\partial
c_n}+\dots+c_{i+1}\frac{\partial}{\partial
c_{i}}+\dots+c_2\frac{\partial}{\partial c_1}]\} \,c_1\,,
\label{odeNsol}
\end{equation}
and it is obvious that (here $m=1, \dots ,n-1$)
\begin{equation}
\frac{\partial^m u(t,\vec{c})}{\partial t^m}={\bf T}_0\exp \{
\int_a^t d\tau\,[f(\tau,\vec{c})\frac{\partial}{\partial
c_n}+\dots+c_{i+1}\frac{\partial}{\partial
c_{i}}+\dots+c_2\frac{\partial}{\partial c_1}]\} \,c_{m+1}\,. \notag
\end{equation}
If we consider $c_1,...,c_n$ as arbitrary constants, the expression
(\ref{odeNsol}) is the general solution of the equation
(\ref{odeN}).

As we have indicated in Example 1 (see Subsection 3.9), the
functions
\begin{equation}
\zeta_i(t,\vec{c})={\bf T}\exp \{ -\int_a^t
d\tau\,[f(\tau,\vec{c})\frac{\partial}{\partial
c_n}+\dots+c_{i+1}\frac{\partial}{\partial
c_{i}}+\dots+c_2\frac{\partial}{\partial c_1}]\} \,c_i\,, \notag
\end{equation}
obey the differential equation
\begin{equation}
\frac{\partial \zeta_i}{\partial t}+f(s,\vec{c})\frac{\partial
\zeta_i}{\partial c_n}+\dots+c_{i+1}\frac{\partial \zeta_i}{\partial
c_{i}}+\dots+c_2\frac{\partial \zeta_i}{\partial c_1}=0\qquad
(\zeta_i|_{t=a}=c_i) \label{odeNzeta}
\end{equation}
and
\[\zeta_i(t,u,\frac{d u}{dt},\dots,\frac{d^{n-1} u}{dt^{n-1}})=c_i\,,\qquad(i=1, \dots ,n).\]
So if we know some, say $k\leq n$, independent solutions of
(\ref{odeNzeta}), then we can eliminate $k$ unknowns from the system
\[\zeta_i(t,u,\frac{d u}{dt},\dots,\frac{d^{n-1} u}{dt^{n-1}})=c_i\,,\qquad(i=1, \dots ,k).\]
and as a result reduce the order of initial problem from $n$ to
$(n-k)$.

In the ``integral-free" form
\begin{equation}
u(t,\vec{c})=[\exp \{ (t-a) \,[f(s,\vec{c})\frac{\partial}{\partial
c_n}+\dots+c_{i+1}\frac{\partial}{\partial
c_{i}}+\dots+c_2\frac{\partial}{\partial
c_1}+\frac{\partial}{\partial s}]\} \,c_1\,]\,|_{s=a}\,, \notag
\end{equation}
the auxiliary function
\begin{equation}
U(t,\vec{c},s)=\exp \{ (t-a) \,[f(s,\vec{c})\frac{\partial}{\partial
c_n}+\dots+c_{i+1}\frac{\partial}{\partial
c_{i}}+\dots+c_2\frac{\partial}{\partial
c_1}+\frac{\partial}{\partial s}]\} \,c_1 \notag
\end{equation}
satisfies the following PDE
\begin{equation}
\frac{\partial U}{\partial t}=f(s,\vec{c})\frac{\partial U}{\partial
c_n}+\dots+c_{i+1}\frac{\partial U}{\partial
c_{i}}+\dots+c_2\frac{\partial U}{\partial c_1}+\frac{\partial
U}{\partial s}\,,\qquad (U|_{t=a}=c_1) \notag
\end{equation}
and
\[u(t,\vec{c})=U(t,\vec{c},s)|_{s=a}\,.\]

Since here we have a larger number of degrees of freedom than in the
case $n=1$, we can obtain many equivalent forms of the formal
solutions like (\ref{Asol3}) using the way described in
the previous section.

\section{Helmholtz equation}

Here we consider the essential features of the operator method for
linear partial differential equations on example of formal solution
of Helmholtz equation under different formulations of the boundary
conditions. The solutions of the Helmholtz equation represent the
(spatial part of) solutions of the wave equation.

Let us consider the Helmholtz equation for an \emph{inhomogeneous}
medium, which has the following form
\begin{equation}
\frac{\partial^2 u(x,y,z)}{\partial
x^2}=-[\Delta_2+\varepsilon(x,y,z)]u(x,y,z)+q(x,y,z) \label{Helm}
\end{equation}
with boundary conditions on $x=a$
\begin{equation}
u(x,y,z)|_{x=a}=\alpha(y,z)\,, \qquad \frac{\partial
u(x,y,z)}{\partial x}|_{x=a}=\beta(y,z)\,,\label{Helmin}
\end{equation}
 where $\Delta_2=\frac{\partial^2
}{\partial y^2}+\frac{\partial^2 }{\partial z^2}$,
$\varepsilon(x,y,z)$ is a function, which takes into account the
wave speed dependence on the space point, and $q(x,y,z)$ is a wave
source.

If we introduce the function
\[S(x,y,z;h,p)=h\,u(x,y,z)+p\,\frac{\partial u(x,y,z)
}{\partial x},\] where $h$ and $p$ are auxiliary real parameters,
then equation (\ref{Helm}) can be transformed into the following form
\[\frac{\partial S(x,y,z;h,p)}{\partial x}=[h\,\frac{\partial
}{\partial p}-[\Delta_2+\varepsilon(x,y,z)]\,p\,\frac{\partial
}{\partial h}]S(x,y,z;h,p)+p\,q(x,y,z)\] with boundary condition on
$x=a$
\[S(x,y,z;h,p)|_{x=a}=h\,\alpha(y,z)+p\,\beta(y,z)\,.\]
Hence the solution of the Helmholtz equation (\ref{Helm}) under
boundary conditions (\ref{Helmin}) is ($x\geq a$)
\begin{align}
&u(x,y,z)=\frac{\partial }{\partial h}\,{\bf T}\exp \{ \int_a^x
d\tau\,[h\,\frac{\partial }{\partial
p}-[\Delta_2+\varepsilon(\tau,y,z)]\,p\,\frac{\partial }{\partial
h}]\} \,[h\,\alpha(y,z)+p\,\beta(y,z)]+\notag\\&\frac{\partial
}{\partial h}\,\int_a^x d\tau\,{\bf T}\exp \{\int_\tau^x
d\xi\,[h\,\frac{\partial }{\partial
p}-[\Delta_2+\varepsilon(\xi,y,z)]\,p\,\frac{\partial }{\partial
h}]\}\, p\,\,q(\tau,y,z)\,.\label{Helmsol}
\end{align}
It is easy to see that $u(x,y,z)$ in (\ref{Helmsol}) does not depend
on auxiliary parameters $h$ and $p$.

There are boundary value problems for Helmholtz equation (which are
more profound from a physical point of view) when one puts certain
requirements on the solution behavior at infinity (at
$r=(x^2+y^2+z^2)^{1/2}\rightarrow\infty$). Since here the
co-ordinate $r$ is selected by boundary conditions it is expedient
to solve this problem in spherical co-ordinates with the initial
supposition that $u$ and its first derivative on $r$ at $r = a$ are
known. At $a\rightarrow\infty$, $u$ and its derivative in a
medium with $Im \, \varepsilon \geq 0$ have to diminish
rapidly enough (there are not wave sources at infinity), hence the first
item of the expression of type (\ref{Helmsol}) goes to zero.
Therefore for $u$ to satisfy the \emph{radiation
conditions} at infinity in a \emph{boundless} inhomogeneous medium
we can find the following operator expression ($Im \, \varepsilon
\geq 0$)
\begin{equation}
u(\vec{r})=-\frac{\partial }{\partial h}\,\int_r^\infty d\zeta\,{\bf
T}_0\exp \{\int_\zeta^r d\xi\,[\frac{h}{\xi^2}\,\frac{\partial
}{\partial
p}-[r^2\,\Delta+\xi^2\varepsilon(\frac{\xi}{r}\,\vec{r})]\,p\,\frac{\partial
}{\partial h}]\}\,
p\,\zeta^2\,q(\frac{\zeta}{r}\,\vec{r}))\,,\label{Helmsol2}
\end{equation}
where vector $\vec{r}=(x,y,z)$ and $\Delta=\frac{\partial^2
}{\partial x^2}+\frac{\partial^2 }{\partial y^2}+\frac{\partial^2
}{\partial z^2}$ is the ordinary Laplace operator.

The Green function for a boundless inhomogeneous medium follows from
(\ref{Helmsol2}) at $q(r) =\delta(\vec{r}-\vec{r}_0)$. By using the
well-known method of images one can find from (\ref{Helmsol2}) the
Green function say for half-space $x>0$ and to get a solution
of a Helmholtz equation (\ref{Helm}) one has to fulfil the radiation
conditions at infinity and a boundary condition on a plane for
example.

\section{The system of linear first-order PDEs}

Let us consider the systems of \emph{linear} PDEs, which are
\emph{first-order} with respect to $x$ and $y$, using as an example the
following two PDEs for \emph{one} function $u(x,y,\vec{\rho})=u$:
\begin{align}
&\frac{\partial u}{\partial x} ={\bf A}(x,y)u\,
,\notag\\&\label{basesys}\\&\frac{\partial u}{\partial y} ={\bf
B}(x,y)u\,, \notag
\end{align}
where $\vec{\rho}$ is a set of parameters say $z_1,...,z_n$, ${\bf
A}(x,y) ={\bf A}(x,y,\vec{\rho})$ and ${\bf B}(x,y) ={\bf
B}(x,y,\vec{\rho})$ are \emph{linear operators}, which do not depend
on $\frac{\partial}{\partial x}$ and $\frac{\partial}{\partial y}$
explicitly.

We can solve the first of them by obtaining
\begin{equation}
u(x,y) ={\bf T}\exp \{\int_a^x d\tau\,{\bf A}(\tau,y)\} \,
\phi(y,\vec{\rho}) , \notag
\end{equation}
where $\phi(y,\vec{\rho})$ is a yet unknown function. Substituting now
this solution into the second equation of the system we have
\begin{equation}
\frac{\partial}{\partial y}{\bf T}\exp \{\int_a^x d\tau\,{\bf
A}(\tau,y)\} \, \phi(y,\vec{\rho}) ={\bf B}(x,y){\bf T}\,\exp
\{\int_a^x d\tau\,{\bf A}(\tau,y)\} \, \phi(y,\vec{\rho}) \notag
\end{equation}
and with the help of (\ref{DT}) we obtain the differential expression:
\begin{align}
&\frac{\partial \phi(y,\vec{\rho})}{\partial y}=[{\bf T}_0\exp
\{-\int_a^x d\tau\,{\bf A}(\tau,y)\} \,{\bf B}(x,y)\,{\bf T}\exp
\{\int_a^x d\tau\,{\bf A}(\tau,y)\}-\notag\\&\int_a^x d\tau\,{\bf
T}_0\exp \{-\int_a^\tau d\xi\,{\bf A}(\xi,y)\} \,\frac{\partial {\bf
A}(\tau,y)}{\partial y}\,{\bf T}\exp \{\int_a^\tau d\xi\,{\bf
A}(\xi,y)\} \,] \, \phi(y,\vec{\rho})\,,  \label{phi}
\end{align}
which have the form of differential equation for an unknown
$\phi(y,\vec{\rho})$ if and only if the right-hand side of
(\ref{phi}) does not depend on $x$, that is when its derivative with
respect to $x$ is equal to zero, which leads to the well-known
\emph{consistency condition}
\begin{equation}
[{\bf A}(x,y),\,{\bf B}(x,y)]+\frac{\partial {\bf A}(x,y)}{\partial
y}-\frac{\partial {\bf B}(x,y)}{\partial x}=0\,.\label{Ccond}
\end{equation}

We can solve with the help of (\ref{opDEsol}) this operator equation
with respect to ${\bf B}(x,y)$, that is rewrite the consistency
condition as
\begin{align}
&{\bf B}(x,y)={\bf T}\exp \{\int_a^x d\tau\,{\bf A}(\tau,y)\}
\,\times\notag \\ &[{\bf B}(a,y)+\int_a^x d\tau \,{\bf T}_0\exp
\{-\int_a^\tau d\xi\,{\bf A}(\xi,y)\} \,\,\frac{\partial {\bf
A}(\tau,y)}{\partial y}\, \,{\bf
T}\exp \{-\int_a^\tau d\xi\,{\bf A}(\xi,y)\}] \,\times\notag \\
&{\bf T}_0\exp \{-\int_a^x d\tau\,{\bf A}(\tau,y)\}, \label{Bsol}
\end{align}
where ${\bf B}(a,y)={\bf B}(x,y)|_{x=a}$ is an arbitrary linear
operator (which does not depend on $\frac{\partial}{\partial x}$ and
$\frac{\partial}{\partial y}$ explicitly).

The formal solution of the equation (\ref{phi}) is
\begin{align}
&\phi(y,\vec{\rho}) ={\bf T}\exp \{\int_b^y d\zeta\,{\bf T}_0\exp
\{-\int_a^x d\tau\,{\bf A}(\tau,\zeta)\} \,{\bf B}(x,\zeta)\,{\bf
T}\exp \{\int_a^x d\tau\,{\bf A}(\tau,\zeta)\} \, -\notag\\&\int_a^x
d\tau\,{\bf T}_0\exp \{-\int_a^\tau d\xi\,{\bf A}(\xi,\zeta)\}
\,\frac{\partial {\bf A}(\tau,\zeta)}{\partial \zeta}\,{\bf T}\exp
\{\int_a^\tau d\xi\,{\bf A}(\xi,\zeta)\} \}\, c(\vec{\rho})\,,
\label{s2}
\end{align}
where $c(\vec{\rho})$ is an arbitrary function. Substituting now
(\ref{Bsol}) into (\ref{s2}) we receive
\[\phi(y,\vec{\rho}) ={\bf T}\exp \{\int_b^y d\zeta\,{\bf B}(a,\zeta)\}\,c(\vec{\rho})\]
and finally the desired solution of considered system
(\ref{basesys}) if (\ref{Ccond}) holds is
\begin{equation}
u(x,y,\vec{\rho}) ={\bf T}\exp \{\int_a^x d\tau\,{\bf
A}(\tau,y,\vec{\rho})\}{\bf T}\exp \{\int_b^y d\zeta\,{\bf
B}(a,\zeta,\vec{\rho})\}\,c(\vec{\rho})\,,\label{spdesol}
\end{equation}
where $c(\vec{\rho})$ has an obvious definition as
$c(\vec{\rho})=u(x,y,\vec{\rho})|_{x=a, y=b}$.

Since here ${\bf A}(x,y,\vec{\rho})$ and ${\bf B}(x,y,\vec{\rho})$ are
\emph{operators}, then (\ref{spdesol}) can represent solutions of
non-trivial systems of PDEs.

\section{Conclusions}
We have presented some ways in solving DEs by the chronological operator
method. Besides linear first-order DEs and systems of such DEs, we have
obtained operator solutions for linear and non-linear ODEs of
arbitrary order.

It is easy to note that the obtained solutions contain differential
operators with respect to arbitrary constants, which represent
initial conditions of the problem. For more complicated problems,
e.g. for non-linear PDEs, the formal solutions will contain
\emph{variational} differential operators. Some examples we have
touched on can be found in \cite{Kosovtsov1} and \cite{Kosovtsov2}.

In conclusion, we believe that we succeeded in demonstrating the
fact that operator forms of DE solutions can be handled analytically
no worse than ordinary functions. In some cases its transformation
properties are more comfortable than, e.g., for some special
functions. \vskip 1cm {\bf Acknowledgments}

The author would like to thank Reece Heineke for a careful reading
of this paper and some actual suggestions.

\end{document}